\documentclass[10pt,journal,compsoc]{IEEEtran}

\usepackage[justification=centering]{caption}
\usepackage{algorithm,algpseudocode}
\usepackage{amsmath}
\usepackage{graphicx}
\usepackage{times}                               
\usepackage{graphicx}
\usepackage{tabularx}
\usepackage{array}
\usepackage{amsmath}
\usepackage{amssymb}
\usepackage{units}
\usepackage{textcomp,xspace}
\usepackage{enumerate}
\usepackage{multirow}
\usepackage{multicol}
\usepackage{subfigure}
\usepackage{color}

\makeatletter
\def\BState{\State\hskip-\ALG@thistlm}
\makeatother

\algdef{SE}[DOWHILE]{Do}{doWhile}{\algorithmicdo}[1]{\algorithmicwhile\ #1}

\ifCLASSINFOpdf

\else

\fi

\begin{document}

\title{Efficient Realization of Householder Transform through Algorithm-Architecture Co-design for Acceleration of QR Factorization}

\author{Farhad Merchant,
        Tarun Vatwani, Anupam Chattopadhyay, ~\IEEEmembership{Senior Member, IEEE,} Soumyendu Raha, \\S K Nandy, ~\IEEEmembership{Senior Member, IEEE,} and Ranjani Narayan
\IEEEcompsocitemizethanks{\IEEEcompsocthanksitem Farhad Merchant, Tarun Vatwani, and Anupam Chattopadhyay are with School of Computer Science and Engineering,
Nanyang Technological University, Singapore\protect\\
E-mail: \{mamirali,tarun,anupam\}@ntu.edu.sg
\IEEEcompsocthanksitem Soumyendu Raha and S K nandy are with Indian Institute of Science, Bangalore \IEEEcompsocthanksitem Ranjani Narayan is with Morphing Machines Pvt. LTd. }
\thanks{Manuscript received April 19, 2005; revised August 26, 2015.}}


\IEEEtitleabstractindextext{%
\begin{abstract}
QR factorization is a ubiquitous operation in many engineering and scientific applications. In this paper, we present efficient realization of Householder Transform (HT) based QR factorization through algorithm-architecture co-design where we achieve performance improvement of 3-90x in-terms of Gflops/watt over state-of-the-art multicore, General Purpose Graphics Processing Units (GPGPUs), Field Programmable Gate Arrays (FPGAs), and ClearSpeed CSX700. Theoretical and experimental analysis of classical HT is performed for opportunities to exhibit higher degree of parallelism where parallelism is quantified as a number of parallel operations per level in the Directed Acyclic Graph (DAG) of the transform. Based on theoretical analysis of classical HT, an opportunity re-arrange computations in the classical HT is identified that results in Modified HT (MHT) where it is shown that MHT exhibits 1.33x times higher parallelism than classical HT. Experiments in off-the-shelf multicore and General Purpose Graphics Processing Units (GPGPUs) for HT and MHT suggest that MHT is capable of achieving slightly better or equal performance compared to classical HT based QR factorization realizations in the optimized software packages for Dense Linear Algebra (DLA). We implement MHT on a customized platform for Dense Linear Algebra (DLA) and show that MHT achieves 1.3x better performance than native implementation of classical HT on the same accelerator. For custom realization of HT and MHT based QR factorization, we also identify macro operations in the DAGs of HT and MHT that are realized on a Reconfigurable Data-path (RDP). We also observe that due to re-arrangement in the computations in MHT, custom realization of MHT is capable of achieving 12\% better performance improvement over multicore and GPGPUs than the performance improvement reported by General Matrix Multiplication (GEMM) over highly tuned DLA software packages for multicore and GPGPUs which is counter-intuitive.
\end{abstract}

}

\maketitle

\IEEEpeerreviewmaketitle

\section{Introduction}
\label{sec:intro}

QR factorization plays pivotal role in computing solution of linear systems of equation, solving linear least square problems, and computing eigenvalues. Such problems arise in navigation to wireless communication systems. For a matrix $A$ of size $m\times n$, QR factorization is given by 

\begin{align}\label{eqn:qrf1}
  A = QR
\end{align}

\noindent where $Q$ is $m\times m$ orthogonal and $R$ is $m\times n$ upper triangle matrix \cite{Golub1}. There are several methods in the literature to perform QR factorization namely Givens Rotation (GR), Householder Transform (HT), Modified Gram-Schmidt (MGS), and Cholesky QR. Following are the two real life application examples, Kalman Filtering (KF) and QR algorithm where QR factorization is used as a tool to solve certain computational problems. 

\noindent {\bf Application 1: Kalman Filtering}

\begin{figure}[!h]
	\centering
	\includegraphics[scale=0.20]{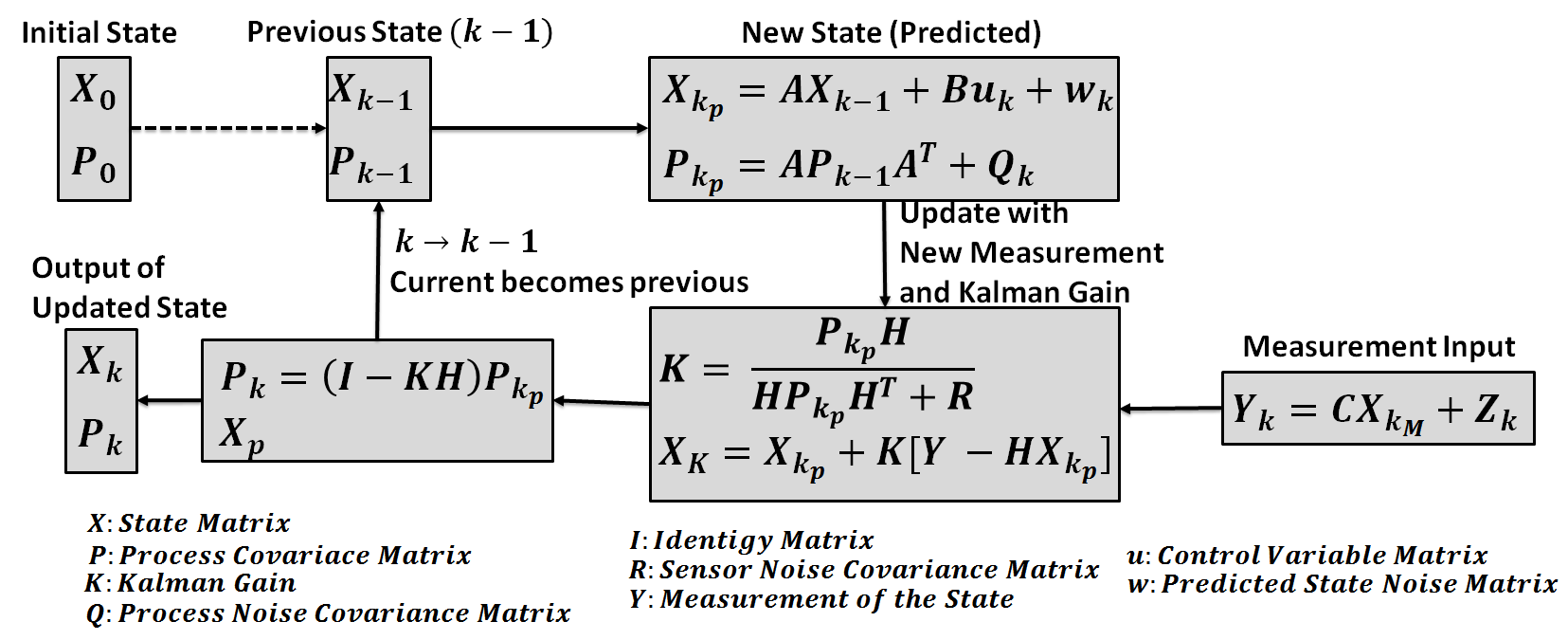}
	\caption{Multi Dimensional Kalman Filter}
	\label{fig:kalman1}
\end{figure}

KF is used in navigation to econometrics since it is capable of filtering out noisy data and at the same time it also facilitates prediction of the next state. A simplistic multi-dimensional KF is shown in the figure \ref{fig:kalman1}. In the KF there is an initial state that contains state matrix and process co-variance matrix. Based on the current state and the previous state, next state is predicted as shown in the figure \ref{fig:kalman1}. Based on the predicted state and co-variance matrix, and measured input, Kalman Gain (KG) is computed and KG is used to predict the next state. Using KG and the predicted state, error in the process is computed. A new state matrix and co-variance matrices are output of the iteration that becomes previous state for the next iteration. 

An iteration of KF requires complex matrix operations ranging from matrix multiplication to computation of numerical stable matrix inverse. One such classical GR based numerical stable approach is presented in \cite{Kal1}. From figure \ref{fig:kalman1}, matrix inverse being the most complex operation in KF, and computation of inverse using QR factorization being a numerical stable process, proposed library based approach is the most suitable for such applications. 


\noindent {\bf Application 2: Eigenvalue Computation}
\begin{algorithm}
\caption{QR Algorithm}
\label{algo:eigen1}
\begin{algorithmic}[1]
\State Let $A^{(0)} = A$
\For K = 1,2,3...
  \State Obtain the factors $Q^{(k)}R^{(k)} = A^{(k-1)}$
  \State Let $A^{(k)} = R^{(k)}Q^{(k)}$
\EndFor
\end{algorithmic}
\end{algorithm}

Computation of eigenvalues is simplified due to QR algorithm where QR algorithm is based on the QR factorization given in equation \ref{eqn:qrf1}. Eigenvalue computation is shown in the algorithm \ref{algo:eigen1}. As a result of QR iteration, eigenvalues appear on the diagonal of the matrix $A$ while columns of the matrix $Q$ are the eigenvectors \cite{qriteration1}. QR algorithm has gained immense popularity for computing eigenvalues and eigenvectors. Some of the examples where eigenvalues and eigenvectors are useful are in communication where eigenvalues and eigenvectors are computed to determine the theoretical limit of the communication medium for the transfer of the information, dimentionality reduction in the principle component analysis for face recognition, and graph clustering \cite{graph_cluster1}\cite{strang1}.

\begin{figure}[!h]
	\centering
	\includegraphics[scale=0.11]{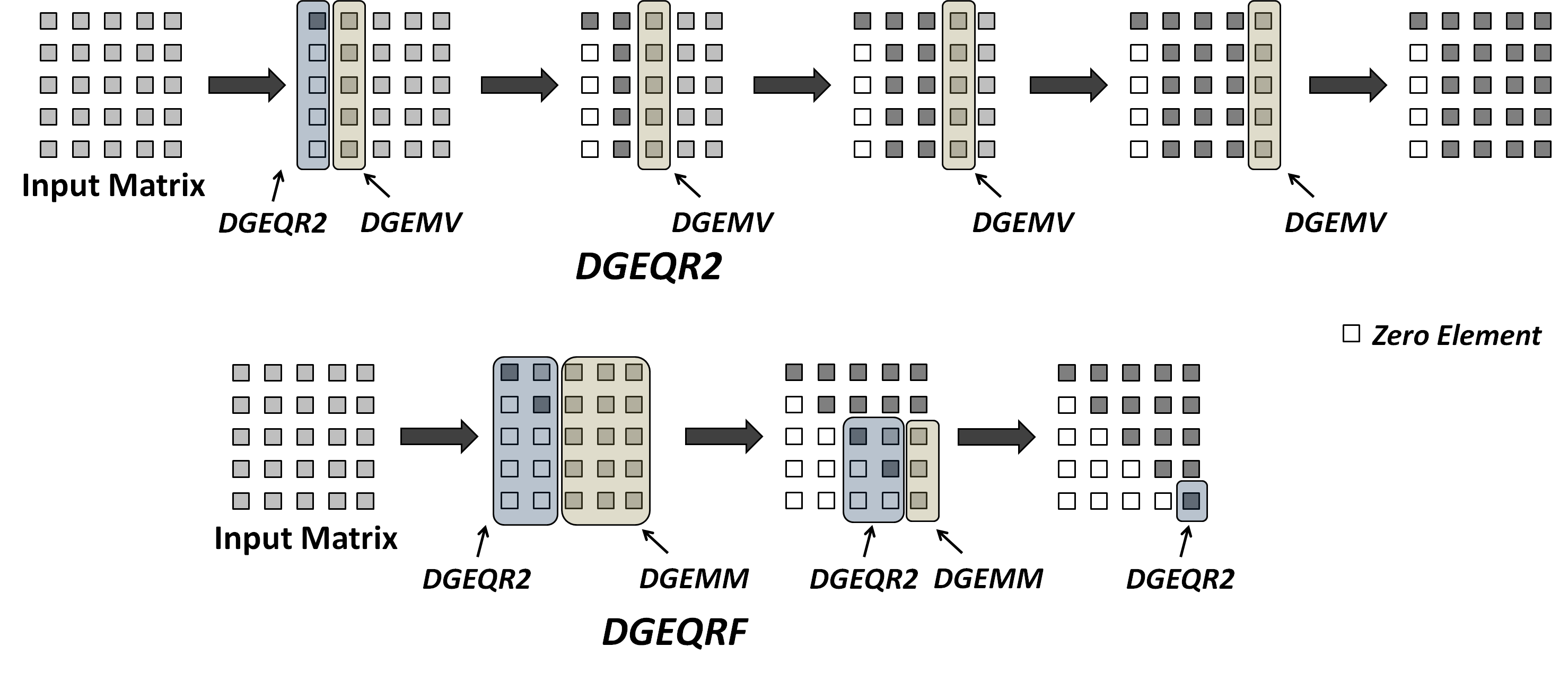}
	\caption{DGEQR2 and DGEQRF Routines}
	\label{fig:dgeqr2_dgeqrf}
\end{figure}

Considering important engineering and scientific applications of QR factorization, it is momentous to accelerate QR factorization. Traditionally, for scalable realization of QR factorization, library based approach is favored due to modularity and efficiency \cite{lapack1}\cite{plasma1}. Graphical representations of HT based QR factorization (XGEQR2) and HT based block QR factorization (XGEQRF) routines in Linear Algebra Package (LAPACK) are shown in figure \ref{fig:dgeqr2_dgeqrf} where X stands for double/single precision version of the routine. It can be observed in the figure \ref{fig:dgeqr2_dgeqrf} that the routine XGEQR2 is dominated by General Matrix-vector (XGEMV) operations while XGEQRF is dominated by General Matrix-matrix (XGEMM) operations. Performance of XGEQRF is observed to be magnitude higher than XGEQR2 due to highly optimized XGEMM operations. XGEMV and XGEMM are part of Basic Linear Algebra Subprograms (BLAS). Typically, performance of LAPACK routines can be measured as a relative performance of BLAS XGEMM since in routines like XGEQRF (QR factorization), XGETRF (LU factorization with partial pivoting), and XPBTRF (Cholesky factorization), XGEMM is dominant and the performance achieved is usually 80\% of the performance achieved by XGEMM for the underlying platform.

Contemporary multicore and General Purpose Graphics Processing Units (GPGPUs) are considered as an ideal platform for efficient realization of BLAS and LAPACK. Multicores are optimized for sequential programs and they are highly efficient in exploiting temporal parallelism exhibited by the routines while GPGPUs are more suitable for the routines that exhibit spatial parallelism \cite{ca1}\cite{ninja1}\cite{blas_fpga1}\cite{gpu1}. Experimentally, none of these platforms are capable of exploiting parallelism that is exhibited by the BLAS and LAPACK routines very efficiently. Moreover, routines in BLAS and LAPACK can be further examined for attaining higher degree of parallelism. We quantify parallelism by depicting routines as a Directed Acyclic Graphs (DAGs) and average number of operation per level ($\beta$) in the DAGs is considered as a measure for fine-grained parallelism exhibited by the routine. Higher $\beta$ means more parallelism in the routine. 

For exploiting spatio-temporal parallelism in BLAS and LAPACK, domain specific customizations are recommended in the platform that is executing these routines \cite{lac1}\cite{lac2}\cite{cgr2}. We choose REDEFINE as a platform for our experiments. REDEFINE is a Coarse-grained Reconfigurable Architecture (CGRA) in which several Tiles are connected through a Network-on-Chip (NoC) \cite{Alle1}\cite{noc1}\cite{cgr1}. Each Tile contains a router for communication and a Compute Element (CE). CEs in REDEFINE can be enhanced with Custom Function Units (CFUs) specifically tailored for domain of interest. CFUs can be Application Specific Integrated Circuits (ASICs), Reconfigurable Data-path (RDP), and/or micro/macro reprogrammable units \cite{hyper1}\cite{Merc1}. In the presented approach we rearrange computations in LAPACK routines vis-\`a-vis amend CFU presented in \cite{Merc1} that can efficiently exploit parallelism exhibited by the modified routine. Thus our approach becomes algorithm-architecture co-design. Considering importance of QR factorization in scientific computing, in this paper we focus on efficient realization of HT based QR factorization through algorithm-architecture co-design. Contributions in this paper are as follows:

\begin{itemize}
 \item Firstly, we discuss evaluation of routines of BLAS and LAPACK on Intel multicore processors and Nvidia GPGPUs. We identify limitations of these machines in exploiting parallelism exhibited by the routines. It is shown that even with highly optimized Dense Linear Algebra (DLA) software packages, the contemporary multicore and GPGPUs are capable of achieving only 0.2-0.3 Gflops/watt
 \item XGEQR2 routine of LAPACK that computes HT based QR factorization is revisited where it is identified that the computations in the classical HT can be re-arranged to exhibit higher degree of parallelism that results in Modified HT (MHT). Empirically, it is shown that realization of MHT (DGEQR2HT) achieves better performance than realization of classical HT (DGEQR2) and better or similar performance compared to DGEQRF in LAPACK while through quantification of parallelism, theoretically it is shown that the parallelism available in the MHT is higher than that of exploited by contemporary multicore and GPGPUs. Realization of MHT on multicore and GPGPU is presented and compared with the state-of-the-art tuned software packages for DLA. Source code of our implementation on multicore and GPGPU is supplied with the exposition
 \item To exploit available parallelism in MHT, we realize MHT on Processing Element (PE) presented in \cite{Merc1}. We adopt methodology presented in \cite{cgr3} and identify macro operations in DAGs of MHT that can be realized on RDP resulting in 1.2-1.3x performance improvement over classical HT realization on the PE. MHT is capable of achieving 99.3\% of the theoretical peak performance achieved by DGEMM in the PE shown in figure \ref{fig:ht_graph_3} which is counter-intuitive as for multicore and GPGPU, the performance achieved by DGEQRF is mostly 80-85\% of the performance achieved by DGEMM. 
 \item Compared to multicore, GPGPU, and ClearSpeed CSX700, 3-80x performance improvement in terms of Gflops/watt is attained. Realization of MHT, outperforms realization of DGEMM as shown in figure \ref{fig:ht_graph_4}. We also show scalability of our solution by attaching PE as a CFU in REDEFINE
\end{itemize}

Due to availability of double precision floating point arithmetic unites like adder, multiplier, square root, and divider, we emphasize on the realization of DGEQR2, and DGEQRF using HT and MHT \cite{fpu2}\cite{fpu3}. Organization of the paper is as follows: In section \ref{sec:rw}, we briefly discuss about REDEFINE and some of the recent realization of QR factorization. In section \ref{sec:mot}, case studies of DGEMM, DGEQR2, and DGEQRF is presented and limitations of the recent multicore and GPGPU in exploiting parallelism are identified. In section \ref{sec:ht}, HT is revisited and MHT is presented and realization of MHT on multicore and GPGPU is presented. In section \ref{sec:cus}, custom realization of HT and MHT is presented in PE. Parallel realization of MHT is also discussed in section \ref{sec:cus}. We summarize our work in \ref{sec:con}.

\noindent {\bf Nomenclature:} 

{\scriptsize

\noindent BLAS\_DGEMM: Legacy realization of DGEMM \\
\noindent LAPACK\_DGEQR2: Legacy realization of HT based QR factorization \\
\noindent LAPACK\_DGEQRF: Legacy realization of HT based block QR factorization\\
\noindent LAPACK\_DGEQR2HT: Realization of MHT based QR factorization\\
\noindent LAPACK\_DGEQRFHT: Realization of MHT based block QR factorization\\
\noindent PLASMA\_ DGEQRF: Legacy realization of HT based tiled QR factorization\\
\noindent PLASMA\_DGEQRFHT: Realization of MHT based block QR factorization\\
\noindent MAGMA\_DGEQR2: Legacy realization of HT based QR factorization\\
\noindent MAGMA\_DGEQRF: Legacy realization of HT based block/tile QR factorization\\
\noindent MAGMA\_DGEQR2HT: Realization of MHT based QR factorization\\
\noindent MAGMA\_DGEQRFHT: Realization of MHT based block/tile QR factorization\\
}

\section{Background and Related Work}
\label{sec:rw}
We revisit REDEFINE micro-architecture briefly and discuss performance of some of the recent DLA computations realized on REDEFINE. Classical HT and its WY representation of HT is also discussed \cite{Golub1}. In the latter part of the section, we discuss some of the recent realization of QR factorization in the literature and their shortcomings. 

\subsection{REDEFINE and DLA on REDEFINE}
\begin{figure}[!ht]
	\begin{centering}
	\includegraphics[scale=0.15]{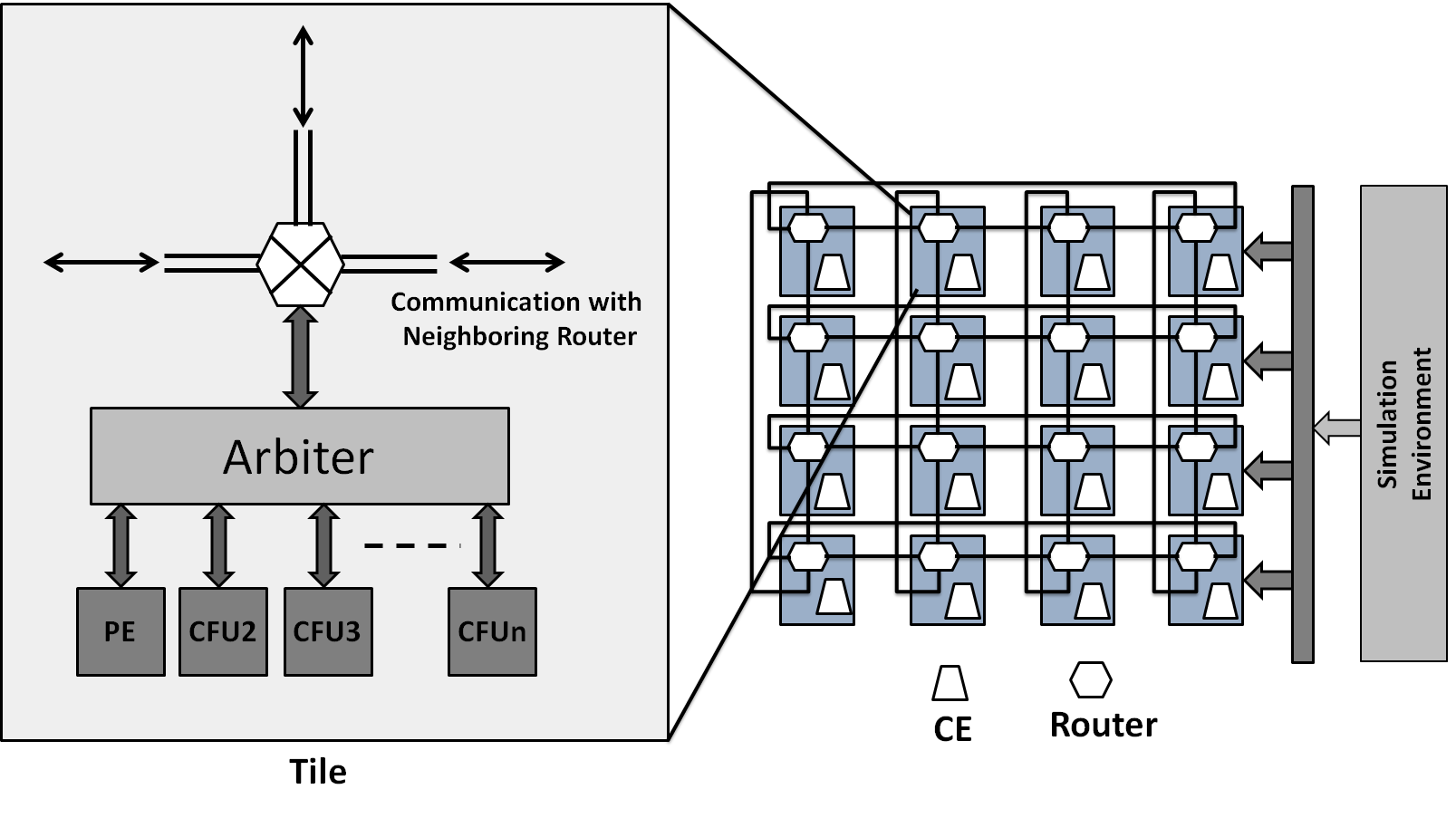}
	\caption{REDEFINE at System Level}
	\label{fig:redefine1}
	\end{centering}
\end{figure}

\begin{figure}[!ht]
	\begin{centering}
	\includegraphics[scale=0.15]{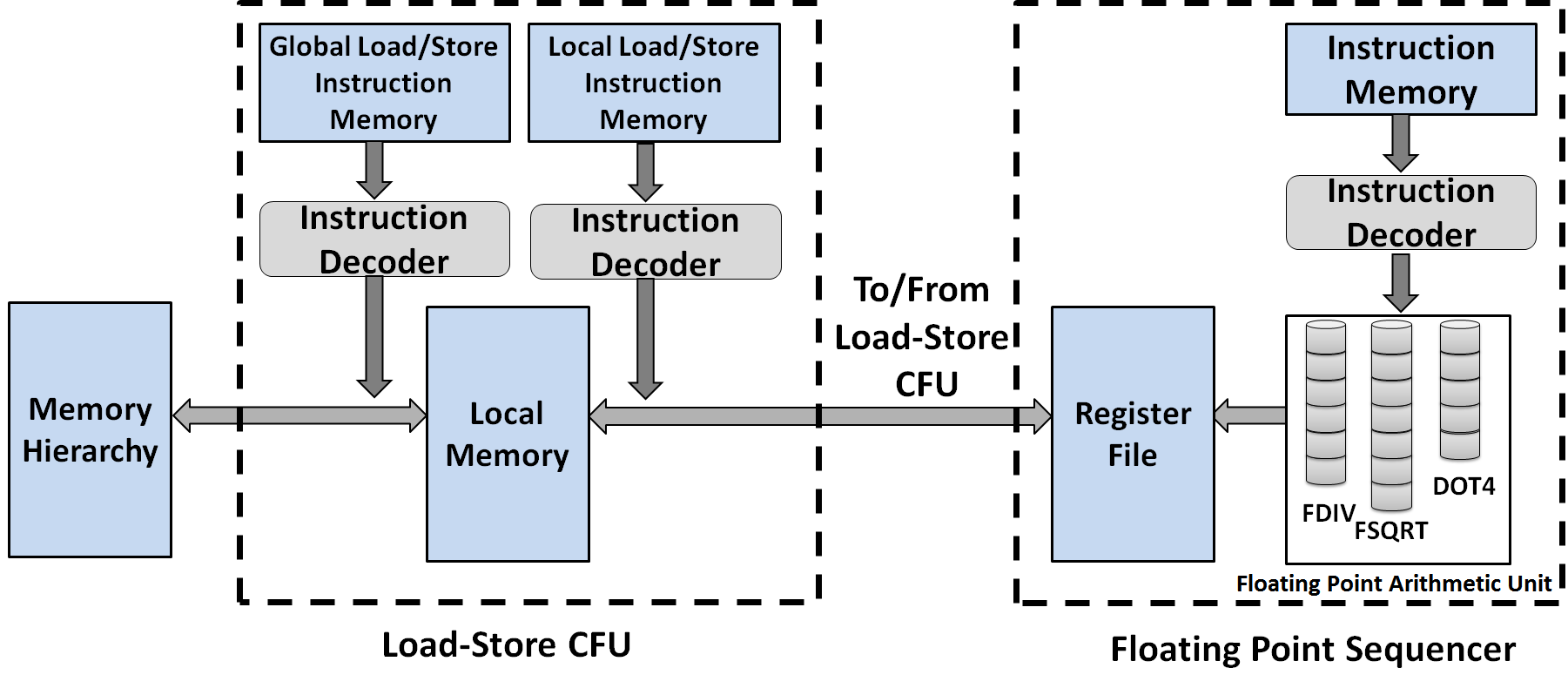}
	\caption{Processing Element Design}
	\label{fig:cfu1}
	\end{centering}
\end{figure}

A system level diagram of REDEFINE is shown in figure \ref{fig:redefine1} where a PE designed for efficient realization of DLA is attached. Micro-architecture of the PE is depicted in figure \ref{fig:cfu1}.

\begin{figure*}
\centering
\subfigure[Percentage of Theoretical Peak of PE attained in DGEMM with Each Architectural Enhancement\label{fig:mm_explo5}]{\includegraphics[scale = 0.21]{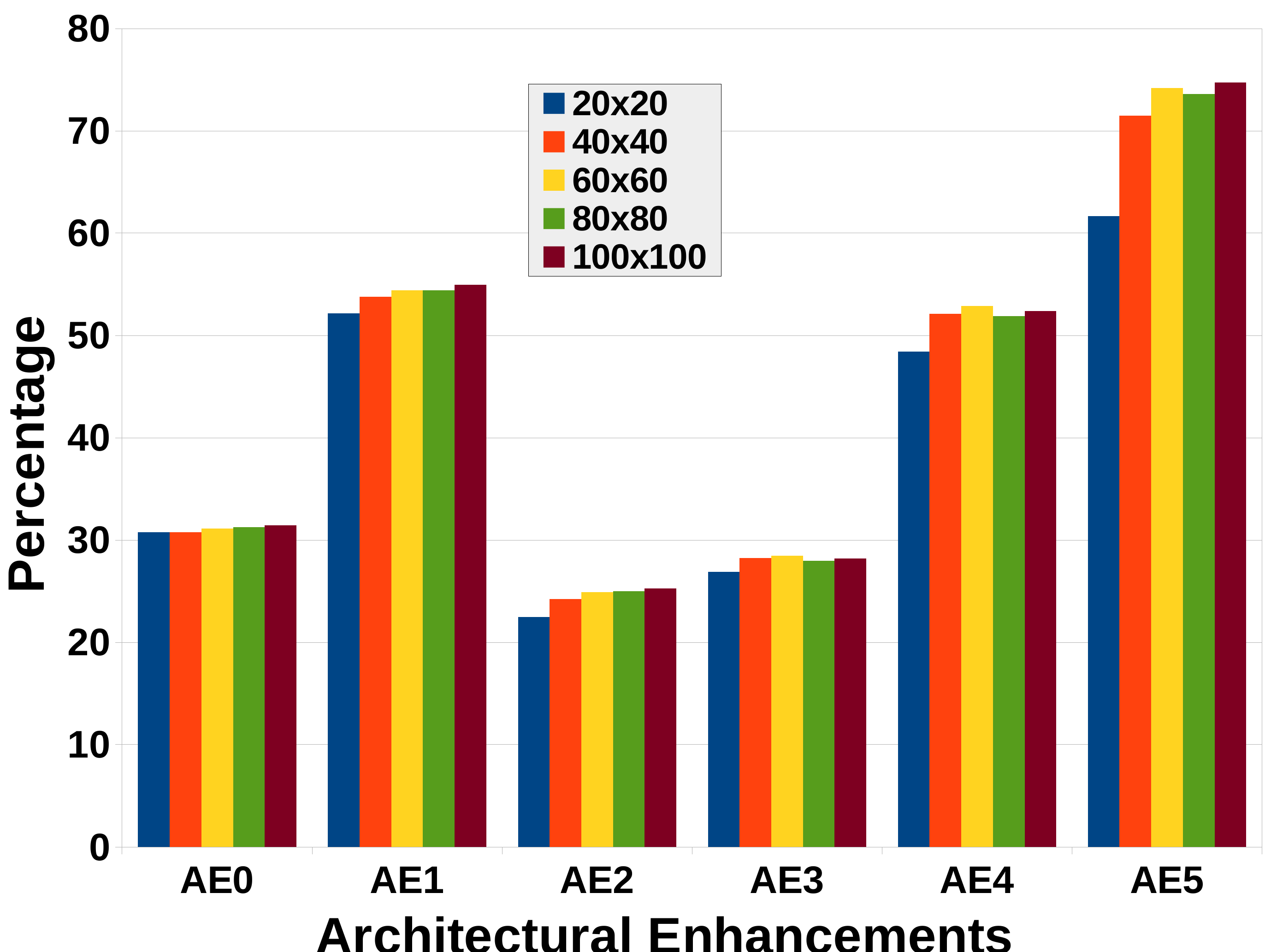}}
\subfigure[Gflops/watt for DGEMM at 0.2GHz, 0.33GHz, 0.95GHz, and 1.81GHz\label{fig:mm_explo8}]{\includegraphics[scale = 0.21]{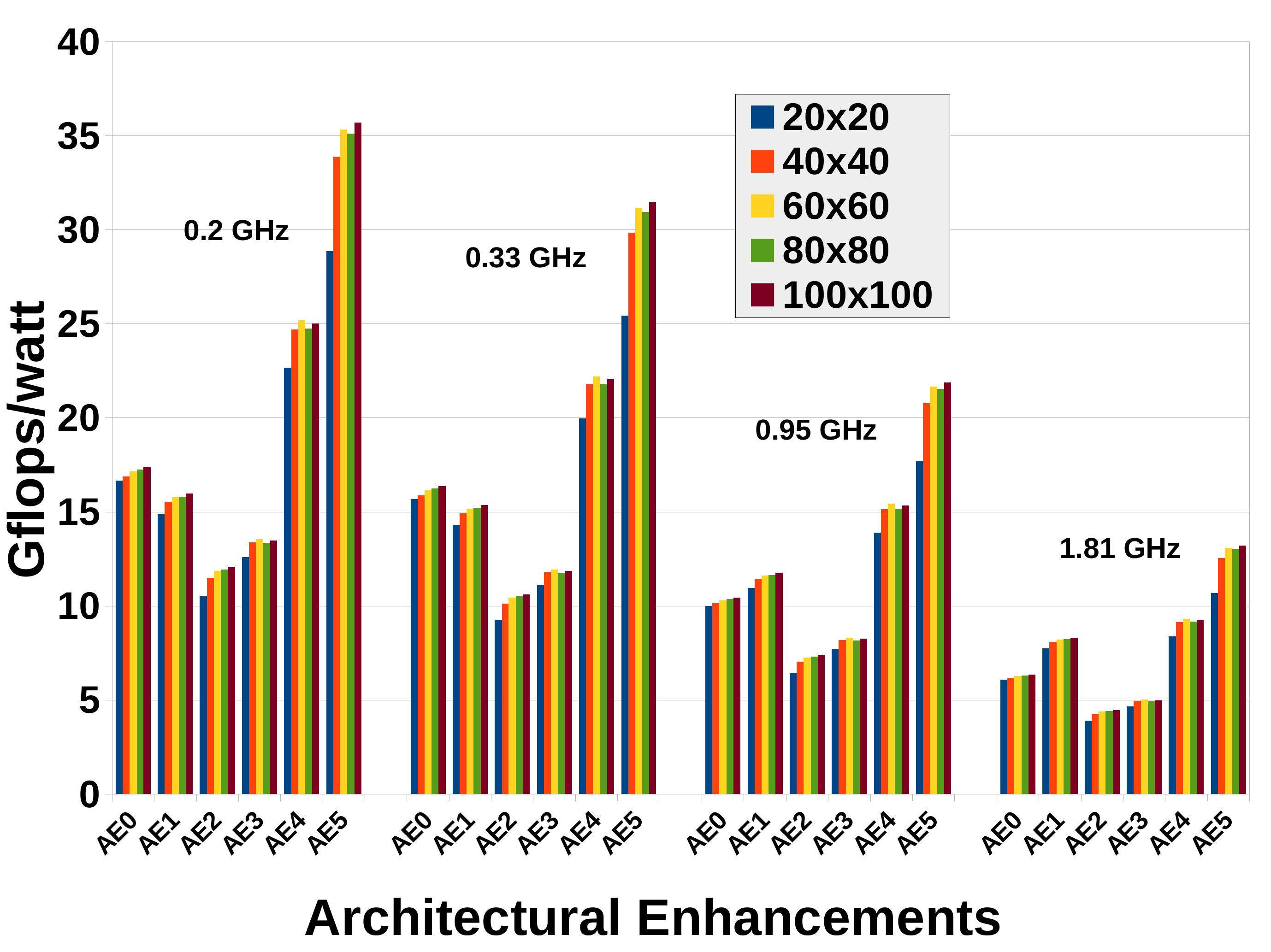}}
\subfigure[Performance Comparison of PE with Other Platforms\label{fig:mm_explo10}]{\includegraphics[scale = 0.21]{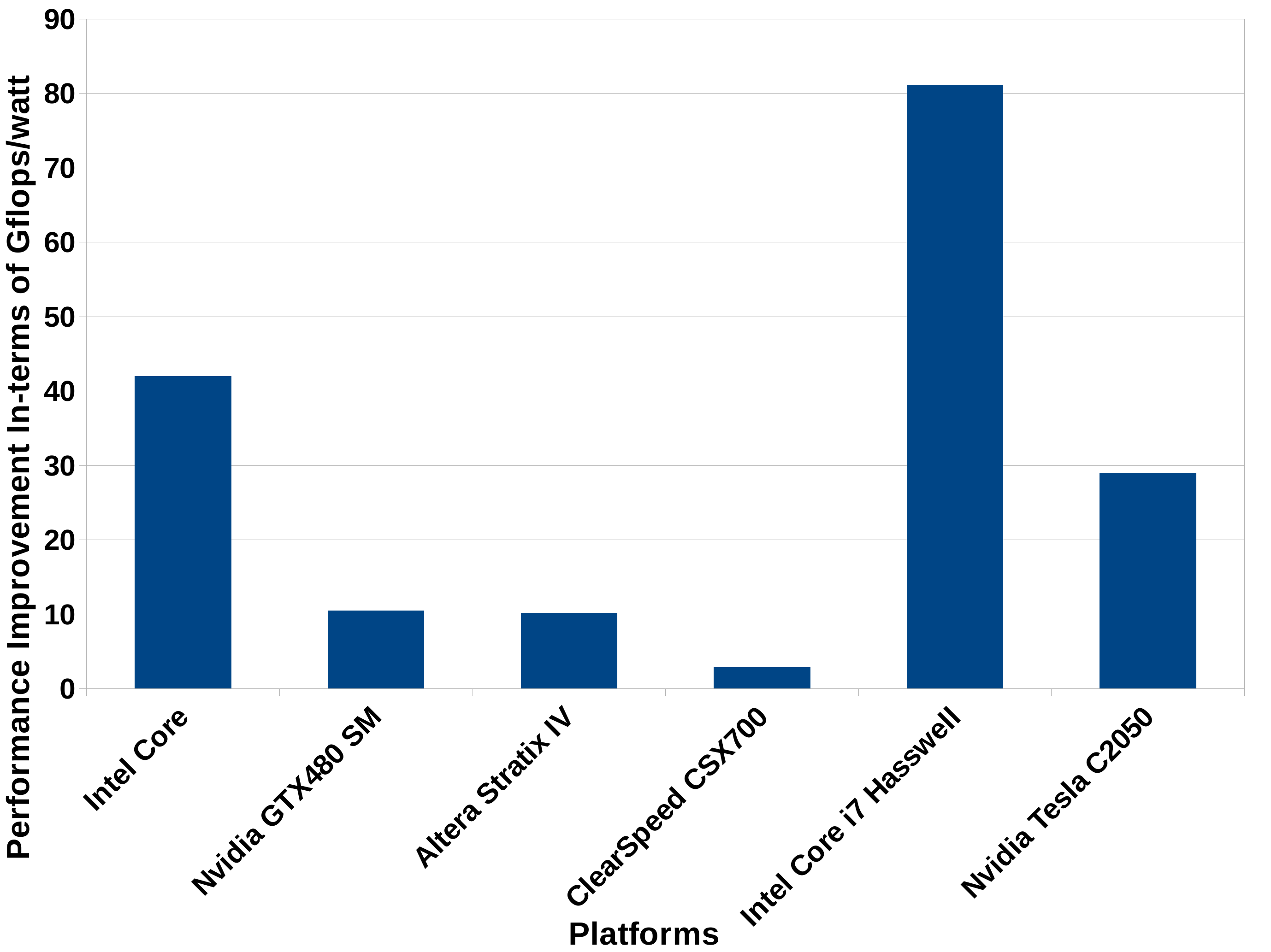}}
\caption{Performance of DGEMM in PE \cite{tpds1} }
\label{fig:dgemm_perf}
\end{figure*}
The PE is taken through several architectural enhancements to improve the performance of the PE and also to ensure maximal overlap of computations and communication \cite{Merc1}. Performance variations in the PE due to architectural enhancements is shown in figure \ref{fig:mm_explo5} while change in the performance in terms of Gflops/watt due to each enhancement is depicted in figure \ref{fig:mm_explo8}. Further details of PE can be found in \cite{Merc1}, and \cite{exp3}. Due to unique features of PE and REDEFINE, we choose PE for our methodology of algorithm-architecture co-design for HT. It can be observed in figure \ref{fig:mm_explo10} that the PE achieves almost 3-140x performance improvement over some of the commercially available multicore, FPGA, and GPGPUs for DGEMM \cite{tpds1}. It is also shown in \cite{tpds1} that when PE used as a CFU in REDEFINE, facilitates scalable parallel realization of BLAS.

\subsection{Classical HT}

In this section, we briefly explain HT and its WY representation. Householder matrix for annihilation of $m-1$ elements in a matrix $A$ of size $m\times n$ is given by 

\begin{align}\label{eqn:p_comp}
P_{m\times m} = I_{m\times m} - 2v_{m\times 1}v^T_{1\times m}
\end{align} 

where $P$ is orthogonal (for real matrices, and Hermitian for complex matrices) and $v$ is householder vector. 

Computations steps for computing householder vector are as follows for matrix $A = \begin{bmatrix} a_{11} & a_{12} & a_{13}  \\ a_{21} & a_{22} & a_{23} \\ a_{31} & a_{32} & a_{33} \end{bmatrix}$:

\begin{align}
\alpha &= -sign(a_{21})\sqrt{\sum_{j=1}^3a^2_{j1}} \nonumber \\
r &= \sqrt{\frac{1}{2}(\alpha^2-a_{11}\alpha)} \nonumber
\end{align}

\begin{figure}[!ht]
	\begin{centering}
	\includegraphics[scale=0.14]{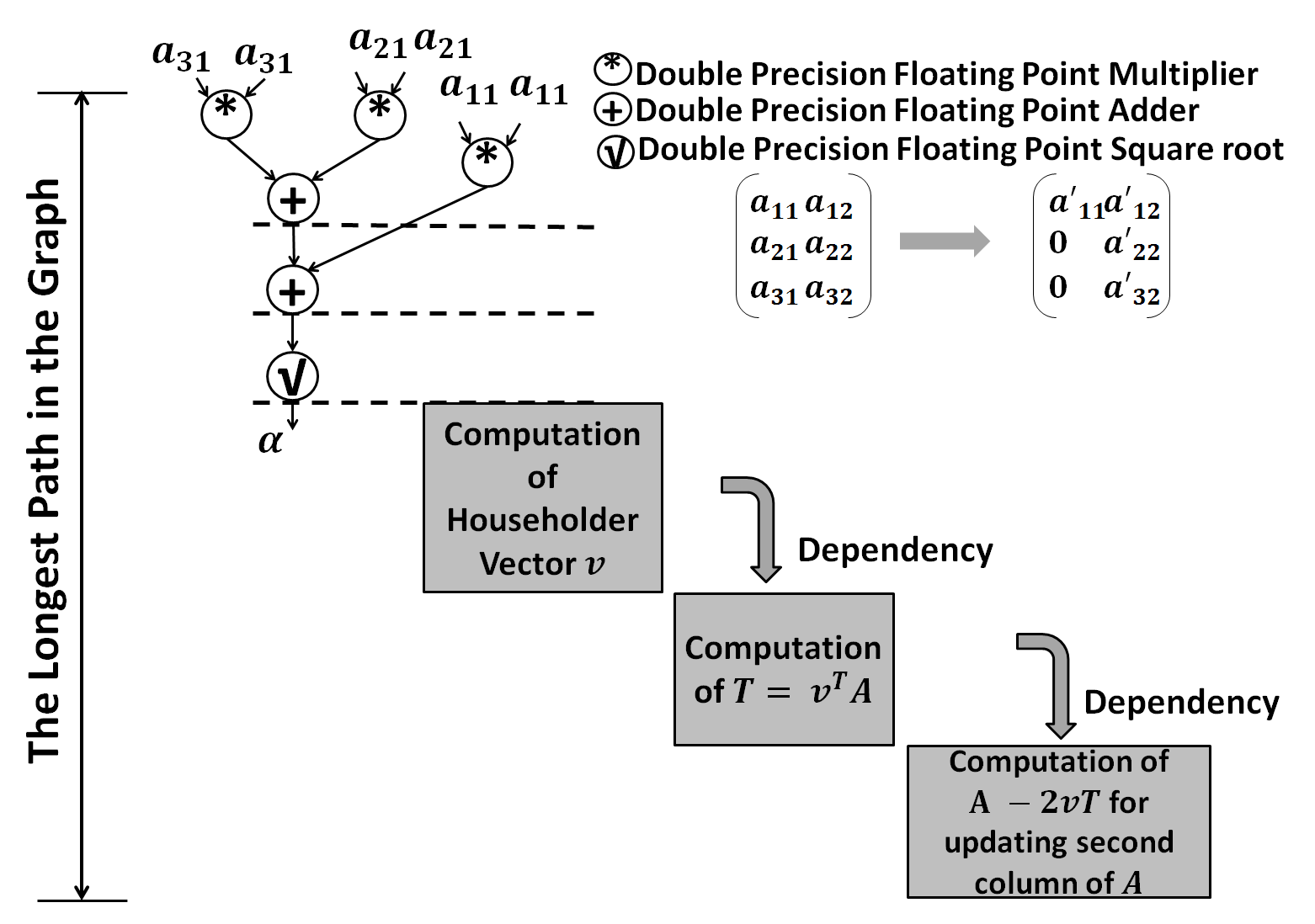}
	\caption{DAGs for Householder Transform}
	\label{fig:dag_ht}
	\end{centering}
\end{figure}

From $\alpha$ and $r$, we can compute $v$ as follows:

\begin{align}
	v = \begin{bmatrix} v_1 \\ v_2 \\ v_3 \end{bmatrix}
\end{align}

where $v_1 = \frac{a_{11} - \alpha}{2r}$, $v_2 = \frac{a_{21}}{2r}$, and $v_3 = \frac{a_{31}}{2r}$. From $v$ vector, $P$ can be computed that annihilates $a_{21}$, and $a_{31}$. $P$ matrix is then multiplied with second and third columns of $A$. In the next iterations, similar procedure is followed to annihilate updated $a_{32}$ by previous iteration.  Directed Acyclic Graphs (DAGs) for annihilation of $a_{31}$ and $a_{21}$ are shown in figure \ref{fig:dag_ht}.  

It can be observed in the figure \ref{fig:dag_ht} that there exist minimal parallelism in computation of $v$ vector. Major source of parallelism is matrix-vector operations and matrix-matrix operations encountered in computation of $P$ matrices. As per Amdahl's law, the performance of the routine is limited by the piece of program that can not be parallelized that is computation of $v$ vectors \cite{amdahl1}.  

\subsection{Related Work}

HT was first presented in \cite{householder1} by Alston S. Householder in 1958 that showed significant improvement in terms of computations over classical Givens Rotation (GR). Since then there were several innovations at algorithmic and technology level that prompted different styles of realizations for HT proposed in \cite{householder1}. First breakthrough arrived with advent of efficient processors with memory hierarchy that induced innovation in efficient realization of Level-3 BLAS \cite{jack7}. Considering efficiency of Level-3 BLAS in the processors at that juncture, there was a successful attempt to realize higher level routines in terms of Level-3 operations. This attempt gave rise to LAPACK, a successor of LINPACK that uses matrix-matrix operations as a basic building block for realization of routines like LU, QR, and Cholesky factorizations \cite{lapack1}. Simultaneously, there was an innovation in HT that resulted in WY-representation of HT presented in \cite{loan1}. The WY-representation proposed in \cite{loan1} is storage efficient and exploits memory hierarchy of the underlying platform more efficiently due to heavy use of Level-3 BLAS operations at comparable operations to its predecessor proposal in \cite{loan2}. Since then WY-representation of HT is preferred due to its computational density and over the years there have been several realizations of HT on contemporary platforms like multicore, GPGPUs, and FPGAs. LAPACK\_DGEQR2 was the first implementation that was dominant in matrix-vector operations (also depicted in figure \ref{fig:dgeqr2_dgeqrf}) while improved version of LAPACK\_DGEQR2 is LAPACK\_DGEQRF that performs block QR factorization based on HT rich in matrix-matrix operations. 

In the recent years, with advent of multicore and GPGPUs, two major packages are developed namely Parallel Linear Algebra Software for Multicore Architectures (PLASMA) for multicore architectures and Matrix Algebra for GPU and Multicore Architectures (MAGMA) for heterogeneous computing environment \cite{plasma1}\cite{magma1}. Corresponding routine realized in PLASMA is PLASMA\_DGEQRF that uses LAPACK\_DGEQR2 and BLAS\_DGEMM along with Queuing and Runtime for Kernels (QUARK) for efficient realization of HT based QR factorization on multicore platforms \cite{quark1}. Similarly, MAGMA uses efficiently realized MAGMA\_BLAS and MAGMA\_DGEQR2 for realization of MAGMA\_DGEQRF. Software stacks of PLASMA and MAGMA are explained in section \ref{sec:plasma_magma_rel}. There have been several attempts for scalable parallel realization of HT based QR factorization on FPGAs \cite{lapack_fpga1}\cite{lapack_fpga2}. The approach presented in \cite{lapack_fpga2} is a multicore based approach emulated on FPGA while LAPACKrc presented in \cite{lapack_fpga1} is a scalable parallel realization of LU, QR and Cholesky factorizations. FPGA based realizations clearly outperform multicore and GPGPU based realizations. A major drawback of FPGA based realization is the energy efficiency of the final solution. It is shown in \cite{tpds1} that CFU tailored for DLA computations clearly outperforms multicore-, GPGPU, and FPGA based solutions and is an ideal platform for our experiments. Surprisingly, multicore, GPGPU, and FPGA based solutions for HT based QR factorization dwell on efficient exploitation of memory hierarchy but none of them focus on rearrangement of computations to expose higher degree of parallelism in HT. In this paper, we present modification to classical HT where we maintain same computation count and memory access of classical HT. We also realize proposed MHT in PLASMA and MAGMA and show marginal improvement in multicore and no improvement in GPGPU.

\section{Case Studies}\label{sec:mot}
We present case study on different available realizations of DGEMM, DGEQR2, and DGEQRF in LAPACK, PLASMA, and MAGMA, and discuss results on multicore and GPGPU platforms.

\begin{figure*}
\centering
\subfigure[CPI Attained in LAPACK\_DGEMM, LAPACK\_DGQR2, and LAPACK\_DGEQRF in Intel Haswell Micro-architectures \label{fig:cpi_gmm_4}]{\includegraphics[scale = 0.21]{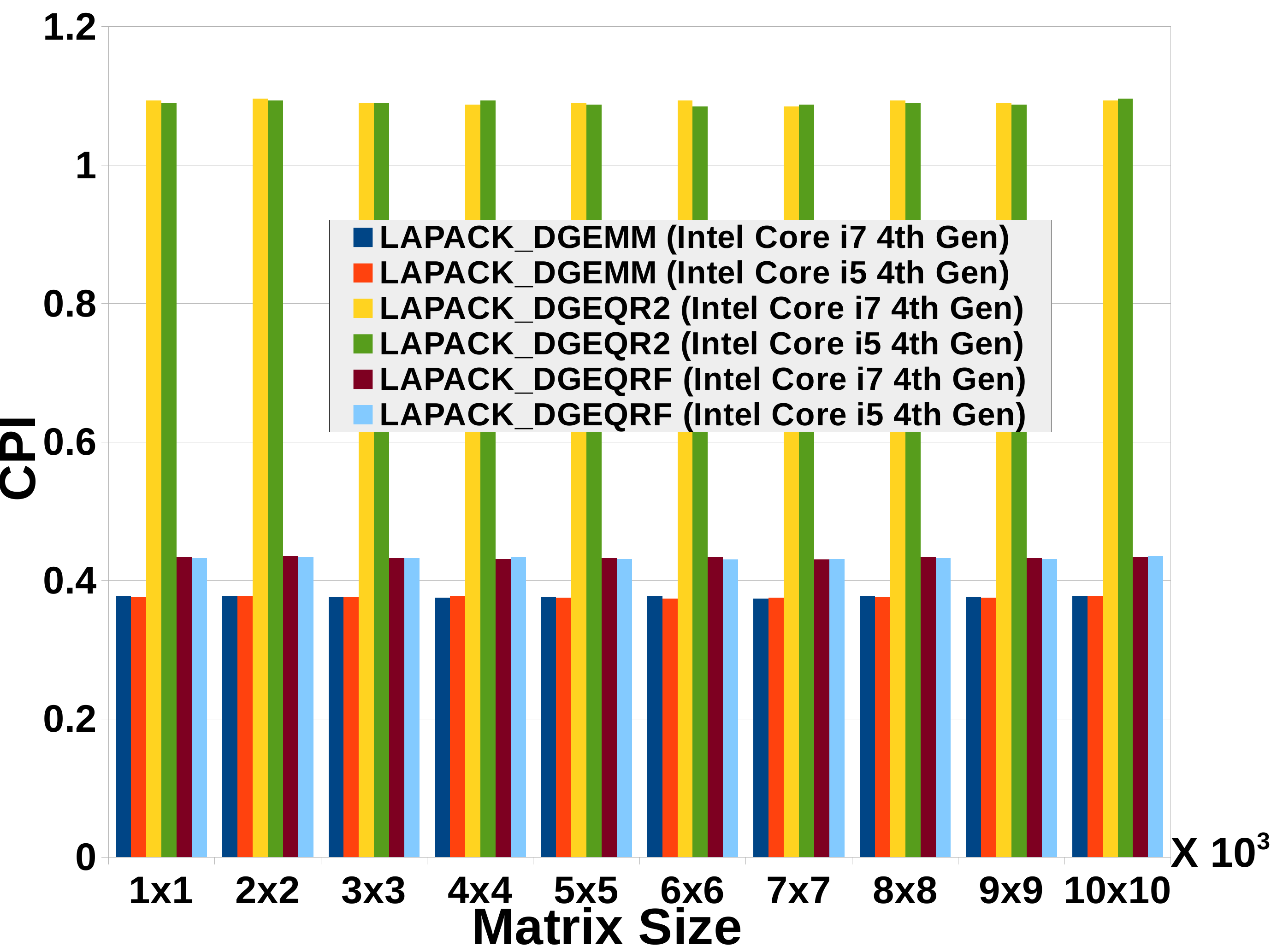}}
\subfigure[Performance of LAPACK\_DGEMM, LAPACK\_DGEQR2, and LAPACK\_DGEQRF In-terms of Gflops in Intel Haswell Micro-architectures \label{fig:cpi_gmm_5}]{\includegraphics[scale = 0.21]{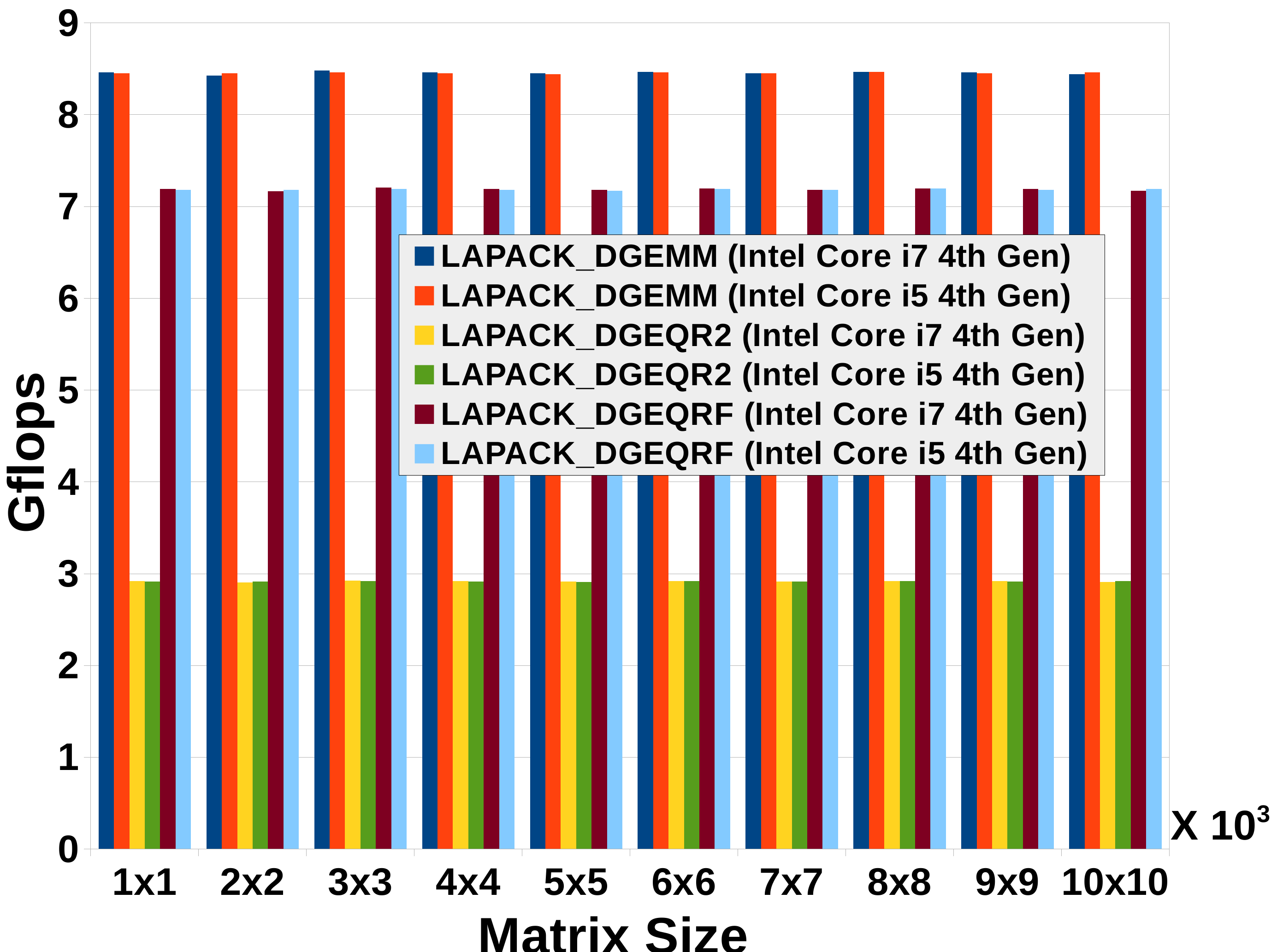}}
\subfigure[Performance of LAPACK/MAGMA\_DGEMM, LAPACK/PLASMA/MAGMA\_DGEQR2, and LAPACK/PLASMA/MAGMA\_DGEQRF in Intel Haswell and Nviida Tesla C2050 Micro-architectures\label{fig:cpi_gmm_6}]{\includegraphics[scale = 0.21]{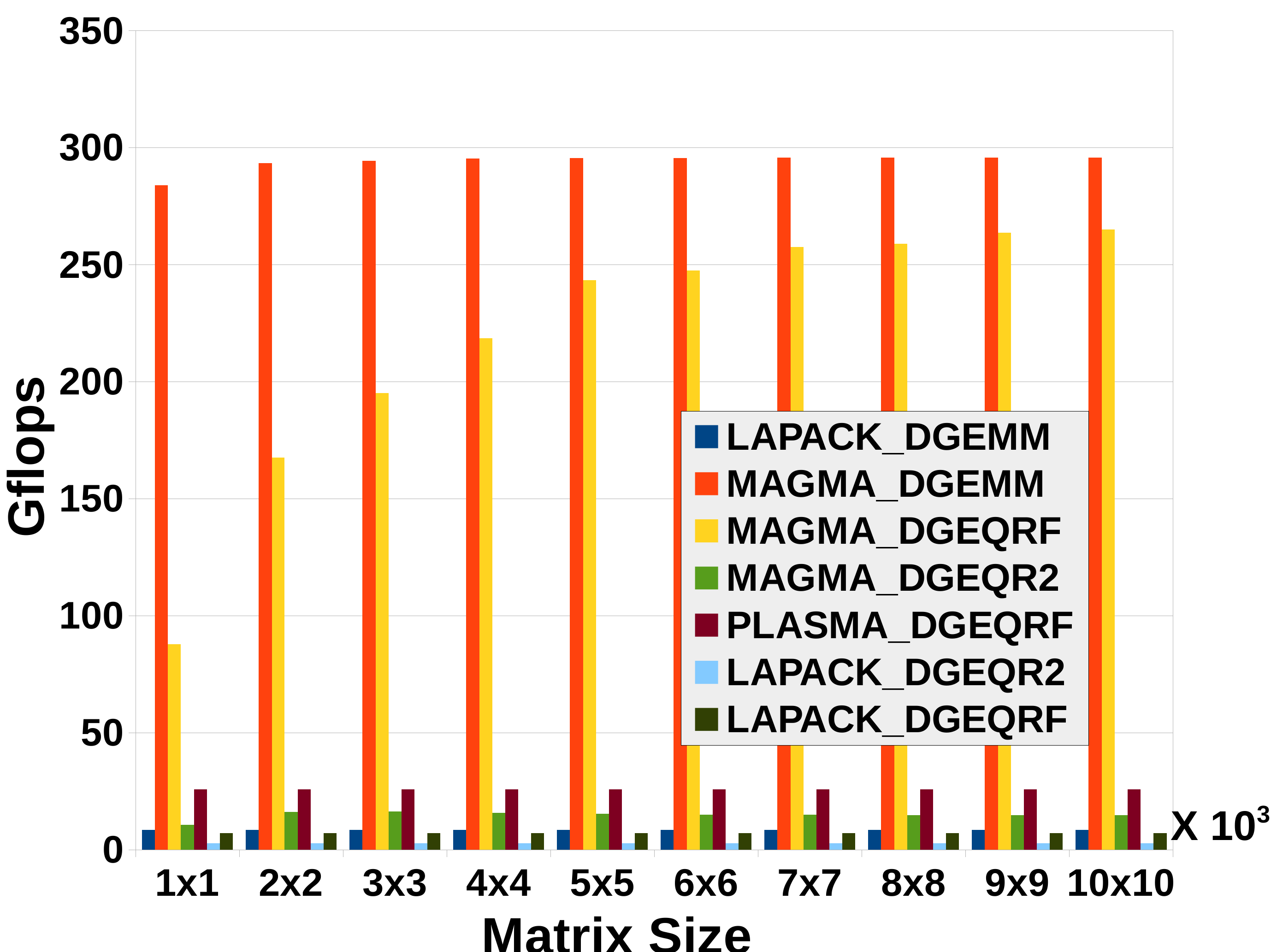}}
\subfigure[Performance Attained In-terms of Theoretical Peak Performance of Underlying Platform\label{fig:cpi_gmm_7}]{\includegraphics[scale = 0.21]{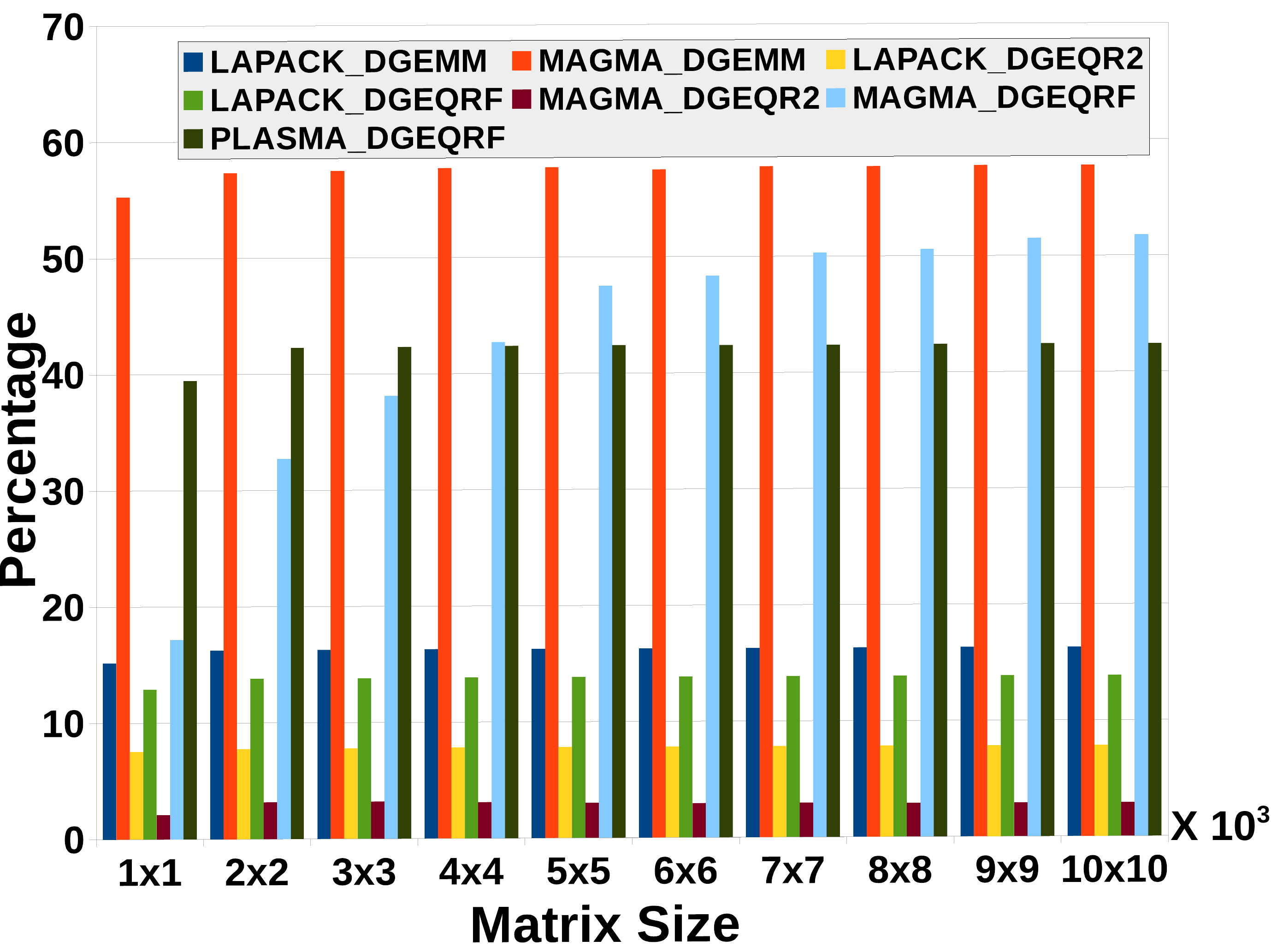}}
\subfigure[Performance Attained In-terms of Gflops/watt for Underlying Platforms\label{fig:cpi_gmm_8}]{\includegraphics[scale = 0.21]{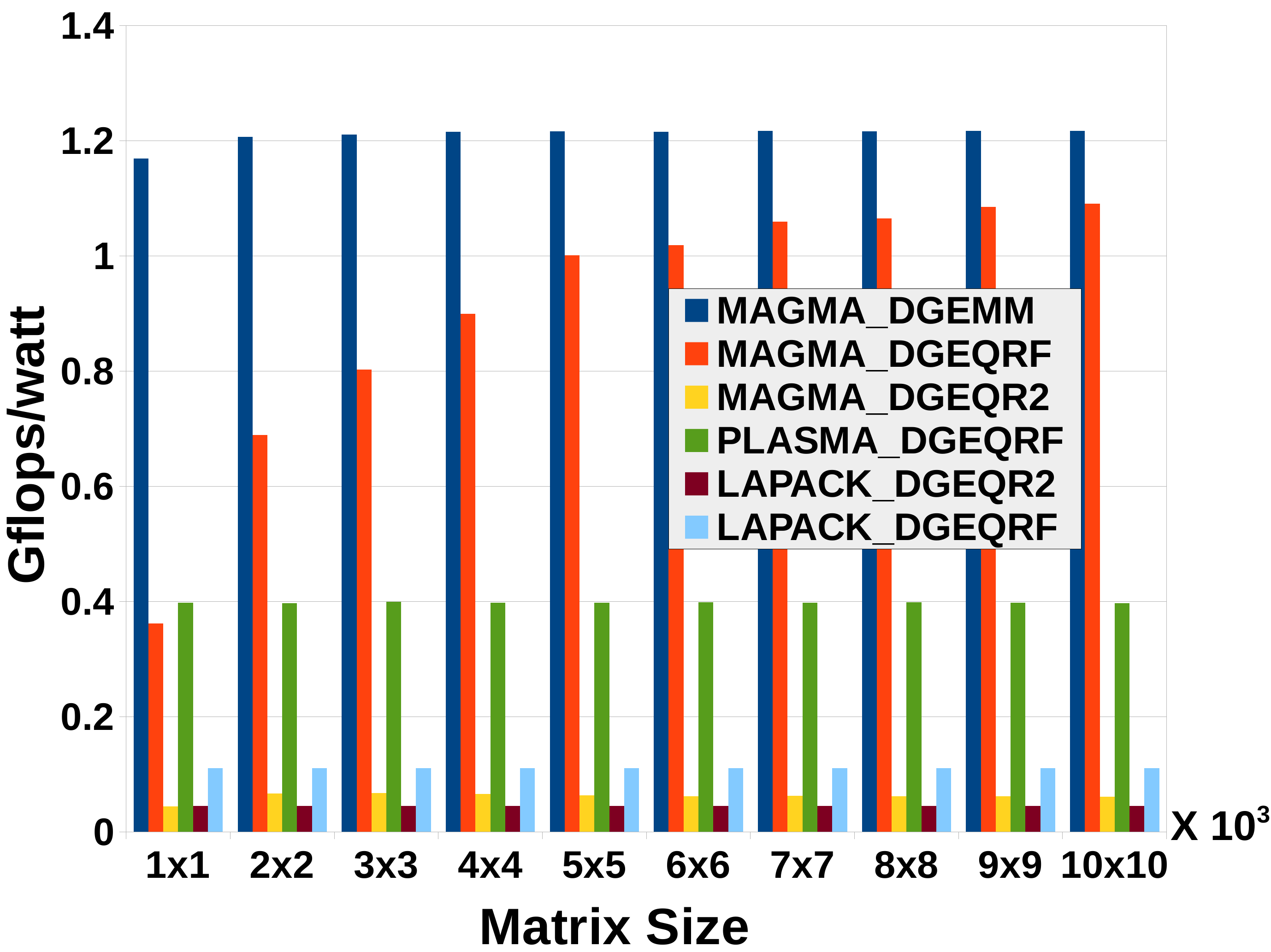}}
\caption{Performance of LAPACK/PLASMA/MAGMA\_DGEQR2, LAPACK/PLASMA/MAGMA\_DGEMM, and LAPACK/PLASMA/MAGM\_DGEQRF on the State-of-the-art Multicore and GPGPU}
\label{fig:dgemm_perf}
\end{figure*}

\subsection{DGEQR2}


\begin{algorithm}
\caption{Pseudo code of DGEQR2}
\label{algo:dgeqr21}
\begin{algorithmic}[1]
\State Allocate memory for input matrix 
\For{$i=1$ to $n$}
  \State Compute Householder vector $v$ 
  \State Compute $P$ where $P = I - 2vv^T$
  \State Update trailing matrix  using DGEMV    
\EndFor
\end{algorithmic}
\end{algorithm}

Pseudo code of DGEQR2 is described in algorithm \ref{algo:dgeqr21}. It can be observed in the pseudo code in the algorithm \ref{algo:dgeqr21} that, it contains three steps, 1) computation of a householder vector for each column 2) computation of householder matrix $P$, and 3) update of trailing matrix using $P = I-2vv^T$ (from equation \ref{eqn:p_comp}). Cycles-per-Instruction (CPI) attained for LAPACK\_DGEQR2 executed on commercially available micro-architectures is shown in figure \ref{fig:cpi_gmm_4}. For our experiments, we use Intel C Compiler (ICC) and Intel Fortran Compiler (IFORT). We also use different compiler switches to improve the performance of LAPACK\_DGEQR2 on Intel micro-architectures. It can be observed in the figure \ref{fig:cpi_gmm_4} that in Intel Core i7 $4^{th}$ Gen machine which is a Haswell micro-architecture, CPI attained saturates at $1.1$. It can be observed in figure \ref{fig:cpi_gmm_5} that attained Gflops saturates at 3 Gflops. Similarly, it can be observed that the percentage of peak performance saturates at 7-8\% of the peak performance in case of Intel Haswell micro-architectures as observed in figure \ref{fig:cpi_gmm_7}. In case when compiler switch $-mavx$ is used that enables use of Advanced Vector Extensions (AVX) instructions, the CPI attained is increased. This behavior is due to AVX instructions that use Fused Multiply Add (FMA). Due to this fact, the CPI reported by VTune\texttrademark can not be considered as a measure of performance for the algorithms and hence we accordingly double the instruction count reported by VTune\texttrademark.

In case of GPGPUs, MAGMA\_DGEQR2 is able to achieve up to 16 Gflops in Tesla C2050 which is 3.1\% of the theoretical peak performance of Tesla C2050 while performance in terms of Gflops/watt is as low as 0.04 Gflops/watt.

\subsection{DGEMM}

Pseudo code for BLAS\_DGEMM routine in BLAS is shown in algorithm \ref{algo:gmm_blas}. It can be observed in the algorithm \ref{algo:gmm_blas} that DGEMM has three nested loops in the algorithm. DGEMM is Level-3 BLAS and it has applications in realization of block algorithms in the DLA software packages since computations in DGEMM are regular in nature and easy to parallelize. CPI attained in BLAS\_DGEMM is 0.37 as shown in figure \ref{fig:cpi_gmm_4} for Intel Haswell micro-architectures. We have used Netlib BLAS for our experiments with platform specific compiler and all the compiler optimizations enabled. BLAS\_DGEMM is able to achieve up to 8.5 Gflops in Intel Haswell micro-architectures with all optimizations as shown in figure \ref{fig:cpi_gmm_5} which is 17\% of the theoretical peak performance of the micro architecture as depicted in figure \ref{fig:cpi_gmm_7}. In case of GPGPU, MAGMA\_DGEMM is able to achieve up to 295 Gflops in Nvidia Tesla C2050 which is 57\% of the peak performance as shown in the figures \ref{fig:cpi_gmm_6} and figure \ref{fig:cpi_gmm_7} respectively. In-terms of Gflops/watt, LAPACK\_DGEMM is capable of attaining 0.12 Gflops/watt in Intel Haswell micro architecture while for Tesla C2050, it is 1.21 Gflops/watt in MAGMA\_DGEMM. 


\begin{algorithm}
\caption{Pseudo code of DGEMM}
\label{algo:gmm_blas}
\begin{algorithmic}[1]
\State Allocate memories for input and output matrices and initialize input matrices
\For{$i=1$ to $m$}
  \For{$j=1$ to $n$}
    \For{$k=1$ to $n$}
      \State C(i,j) = A(i,k)B(k,j) + C(i,j)
     \EndFor
  \EndFor
\EndFor
\end{algorithmic}
\end{algorithm}


\subsection{DGEQRF}


\begin{algorithm}
\caption{Pseudo code of DGEQRF}
\label{algo:dgeqrf1}
\begin{algorithmic}[1]
\State Allocate memories for input matrix
\For{$i=1$ to $n$}
  \State Compute Householder vectors for block column $m\times k$
  \State Compute $P$ matrix where $P$ is Computed using Householder vectors 
  \State Update trailing matrix using DGEMM 
\EndFor
\end{algorithmic}
\end{algorithm}

Pseudo code for DGEQRF routine is shown in algorithm \ref{algo:dgeqrf1}. In terms of computations, there is no difference between algorithms \ref{algo:dgeqr21}, and \ref{algo:dgeqrf1}. In a single core implementation, LAPACK\_DGEQRF is observed to be 2-3x faster than LAPACK\_DGEQR2. The major source of efficiency in LAPACK\_DGEQRF is efficient utilization of processor memory hierarchy and BLAS\_DGEMM routine which is a compute bound operation \cite{rob3}\cite{jack1}. CPI attained in LAPACK\_DGEQRF is 0.43 as shown in figure \ref{fig:cpi_gmm_4} which is much lower than the CPI attained by LAPACK\_DGEQR2. In-terms of Gflops, LAPACK\_DGEQRF is 2-3x better than LAPACK\_DGEQR2 as shown in figure \ref{fig:cpi_gmm_5} while the performance attained by LAPACK\_DGEQRF is 85\% of the performance attained by LAPACK\_DGEMM. LAPACK\_DGEQRF achieves 6-7 Gflops in Intel Haswell micro-architecture as shown in figure \ref{fig:cpi_gmm_5}. In Nvidia Tesla C2050, MAGMA\_DGEQRF is able to achieve up to 265 Gflops as shown in figure \ref{fig:cpi_gmm_6} which is 51.4 \% of theoretical peak performance of Nvidia Tesla C2050 as shown in the figure \ref{fig:cpi_gmm_7} which is 90.5\% of the performance attained by MAGMA\_DGEMM. In MAGMA\_DGEQR2, performance attained in terms of Gflops/watt is as low as 0.05 Gflops/watt while for MAGMA\_DGEMM and MAGMA\_DGEQRF it is 1.21 Gflops/watt and 1.09 Gflops/watt respectively in Nvidia Tesla C2050 as shown in figure \ref{fig:cpi_gmm_8}. In case of PLASMA\_DGEQRF, the performance attained is 0.39 Gflops/watt while running PLASMA\_DGEQRF for four cores. 


Based on our empirical case studies of DGEQR2, DGEMM, and DGEQRF, we make following observations. 

\begin{itemize}
	\item Due to presence of bandwidth bound operations like DGEMV in DGEQR2. the performance of DGEQR2 is not satisfactory in Intel or Nvidia micro-architectures. 
	\item Despite presence of compute bound operations like DGEMM in LAPACK\_DGEQRF, PLASMA\_DGEQRF, and MAGMA\_DGEQRF, these routines are able to achieve only 8-16\% of the peak Gflops in Intel Haswell and 51.5\% of the peak Gflops in Nvidia Tesla C2050 respectively. 
	\item Performance achieved by DGEQR2, DGEMM, and DGEQRF in Intel Haswell and Nvidia C2050 is as low as 0.05-1.23 Gflops/watts
\end{itemize}

Based on above observations, we see a scope in optimization of DGEQR2 and furthermore improvement in DGEQRF routines in LAPACK. In the next section, we continue our quest for optimizations of these routines for commercially available micro-architectures. 

\section{Modified Householder Transform}\label{sec:ht}
In this section, we revisit the classical HT described in algorithm \ref{algo:dgeqr21} and look for further tuning of the algorithm assuming infinite memory bandwidth and infinite number of arithmetic units required for computing $R$ matrix. It can be observed in the algorithm \ref{algo:dgeqr21} that the computations of Householder vector is Level-1 BLAS operation and there is no further scope for optimization in computation of Householder vector. The computations that are dependent on the computation of Householder vector are computation of $P$ matrix which is a Householder matrix and trailing matrix update of the input matrix $A$. In classical HT, the trailing matrix update is performed by pre-multiplying matrix $A$ with the Householder matrix $P$ as shown in equations \ref{eqn:ht1} and \ref{eqn:ht2}.
\begin{align}
	P &= I - 2vv^T \label{eqn:ht1} \\
	PA &= A - 2vv^TA \label{eqn:ht1}
\end{align}


\begin{algorithm}
\caption{Pseudo code of Householder Transform}
\label{algo:dgeqr2_1}
\begin{algorithmic}[1]
\State Allocate memories for input matrix
\For{$i=1$ to $n$}
  \State Compute Householder vectors for block column $m\times k$
  \State Compute $P$ matrix where $P = I - 2vv^T$
  \State Compute $PA$ where $PA = A - 2vv^TA$ 
\EndFor
\end{algorithmic}
\end{algorithm}

Equation \ref{eqn:ht1} in algorithm form is shown in \ref{algo:dgeqr2_1}. It can be observed in the algorithm \ref{algo:dgeqr2_1} that the computation of $2vv^TA$ and computation of $A-2vv^TA$ can be merged. Routine where we merge these two loops is shown in algorithm \ref{algo:dgeqr2ht}.


\begin{algorithm}
\caption{Pseudo code of Modified Householder Transform}
\label{algo:dgeqr2ht}
\begin{algorithmic}[1]
\State Allocate memories for input matrix
\For{$i=1$ to $n$}
  \State Compute Householder vectors for block column $m\times k$
  \State Compute $PA$ where $PA = A - 2vv^TA$
\EndFor
\end{algorithmic}
\end{algorithm}

\begin{figure}[!ht]
	\begin{centering}
	\includegraphics[scale=0.15]{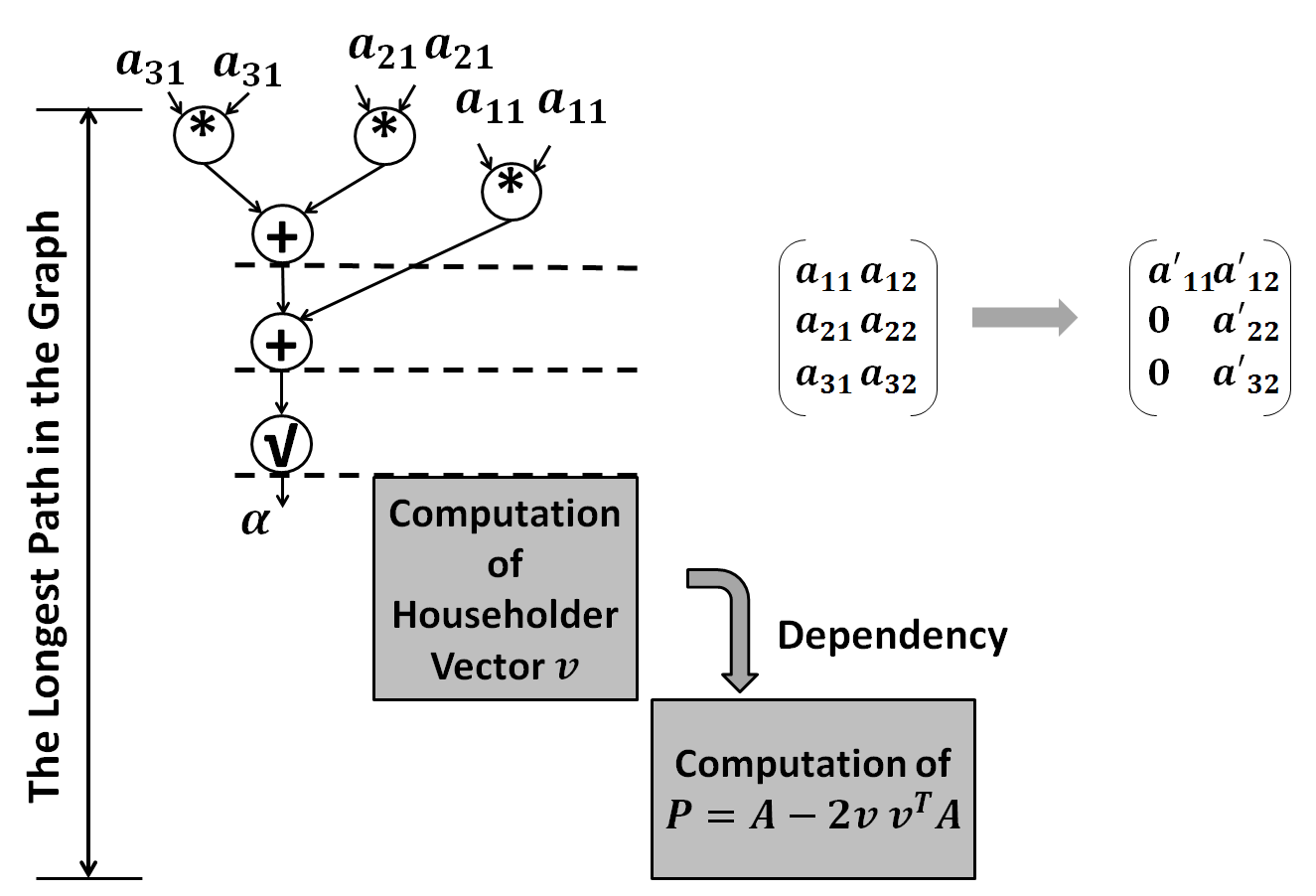}
	\caption{DAGs of MHT}
	\label{fig:dag_dgeqr2ht}
	\end{centering}
\end{figure}

MHT for $3\times 3$ matrix is shown in figure \ref{fig:dag_dgeqr2ht}. It can be observed from the figure \ref{fig:dag_ht} and \ref{fig:dag_dgeqr2ht} that due to fusing of the inner and intermediate loops in MHT, the depth of the graph decreases. It can also be observed that there are more operations per level in the graph. If we take number of operations per level in the DAG of HT as $\beta$ then average $\beta$ is given by equation \ref{eqn:beta_1}.

\begin{align}\label{eqn:beta_1}
	\beta = \frac{\text{Total Number of Computations in the routine}}{\text{Number of Levels in the DAG of the routine}}
\end{align}

\begin{figure}[!ht]
	\begin{centering}
	\includegraphics[scale=0.22]{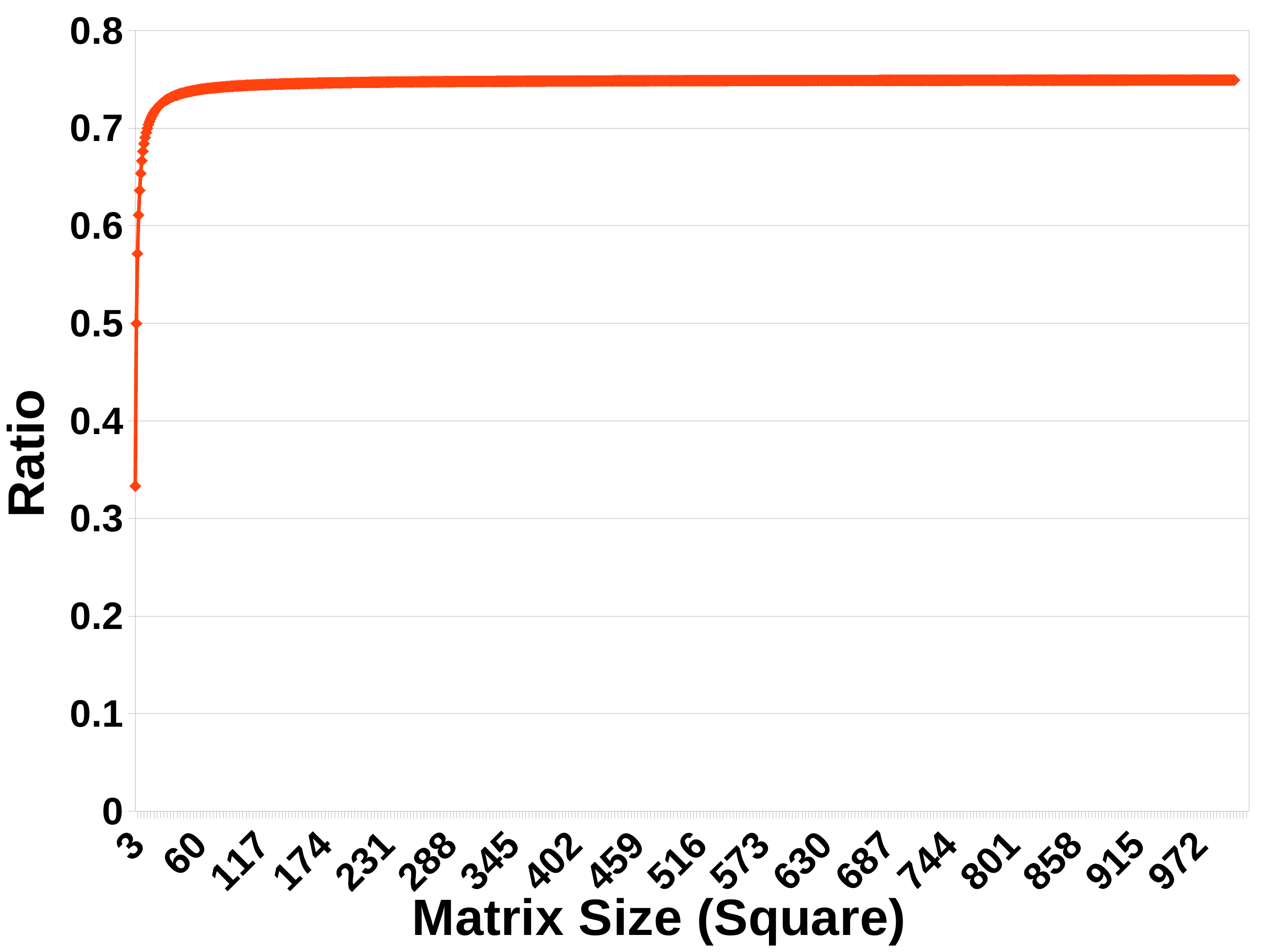}
	\caption{Ratio}
	\label{fig:ratio1}
	\end{centering}
\end{figure}

In case of MHT, since there is a decrease in the number of levels in the DAG of the algorithm, $\beta_{MHT} > \beta_{HT}$ where $\beta_{MHT}$ is $\beta$ for MHT and $\beta_{HT}$ is $\beta$ for HT. Quantifying $\beta$ for HT and MHT,

\begin{align}
 \beta_{HT} &= \frac{\text{Number of Operations in HT}}{\text{Number of Levels in DAG of HT}} \\
 \beta_{MHT} &= \frac{\text{Number of Operations in MHT}}{\text{Number of Levels in DAG of MHT}}
\end{align}

Considering ratio of $\beta_{HT}$ and $\beta_{MHT}$,

\begin{align} \label{eqn:theta1}
 \theta &= \frac{\beta_{HT}}{\beta_{MHT}} \\
	&= \frac{\text{Number of Levels in DAG of MHT}}{\text{Number of Levels in DAG of HT}}
\end{align}

The parameter $\theta$ in equation \ref{eqn:theta1} is ratio of quantified parallelism in HT and MHT respectively. For our analysis $\theta$ is independent of the computations since there is no change in the computations and it is also independent of communication since the communication pattern remain identical in HT and MHT. As value of $\theta$ decreases the parallelism in MHT is more. For HT and MHT, $\theta$ saturates at 0.749 as shown in figure \ref{fig:ratio1}. The method used here to quantify the parallelism in the routine is simple since the computations are regular in nature. For complex algorithms, method described in \cite{quant1} can be used. To support proposition of improvement in the parallelism in MHT, we experiment on several commercially available multicore and GPUs. For our experiments on multicore, we have used LAPACK available in the Netlib with vendor specific optimizations and we have ensured to use LAPACK program semantics while realizing and integrating realization of MHT in LAPACK \cite{lapack1}. For GPGPUs we have used highly tuned MAGMA package and ensured that the program semantics of MAGMA\_DGEQR2HT confirms with MAGMA semantics for ease of integration with the package \cite{magma1}.

\subsection{Realization on Multicore and GPGPU} \label{sec:plasma_magma_rel}

\begin{figure*}
\centering
\subfigure[PLASMA Software Stack \cite{plasma1}\label{fig:plasma_soft}]{\includegraphics[scale = 0.20]{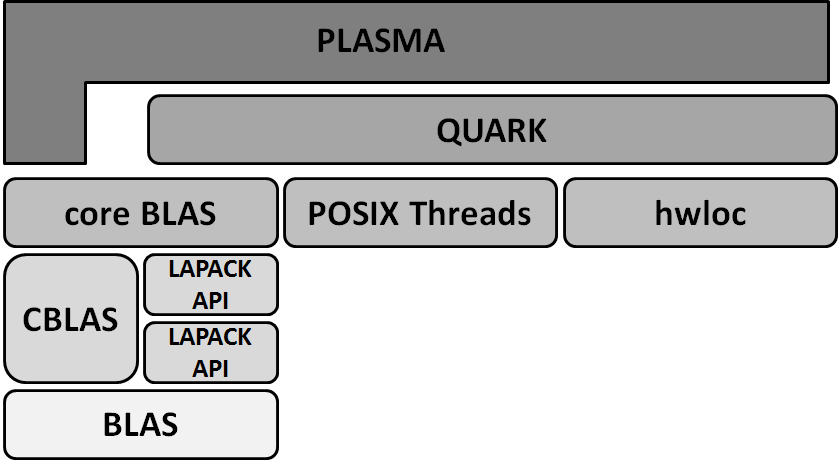}}
\subfigure[MAGMA Software Stack \cite{magma1}\label{fig:magma_soft}]{\includegraphics[scale = 0.20]{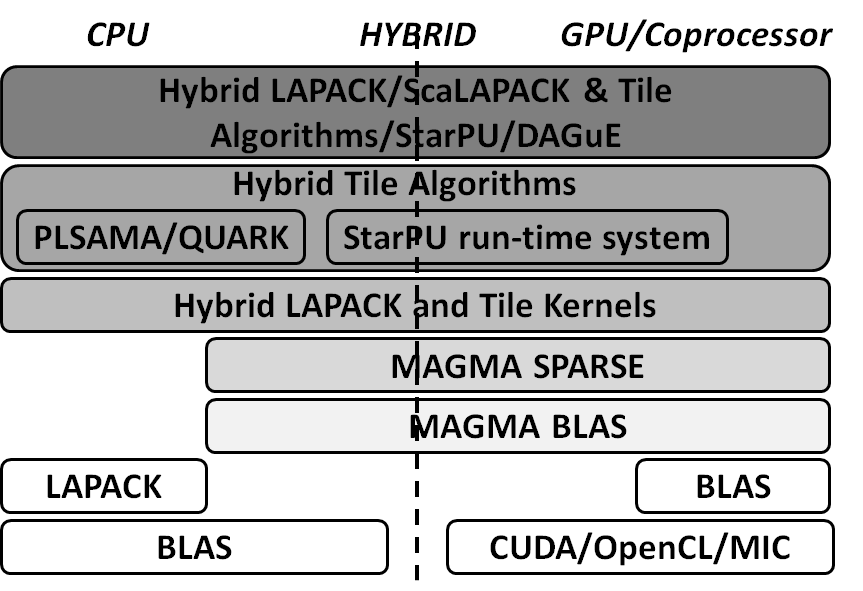}}
\caption{PLASMA and MAGMA Software Stacks}
\label{fig:dgemm_perf}
\end{figure*}

\begin{figure*}
\centering
\subfigure[Performance Comparison of LAPACK\_DGEQR2, LAPACK\_DGEQRF, LAPACK\_DGEQR2HT, and LAPACK\_DGEQRFHT on Intel and AMD Micro-architectures (from \cite{exp3})\label{fig:cpi_dgeqr2ht}]{\includegraphics[scale = 0.21]{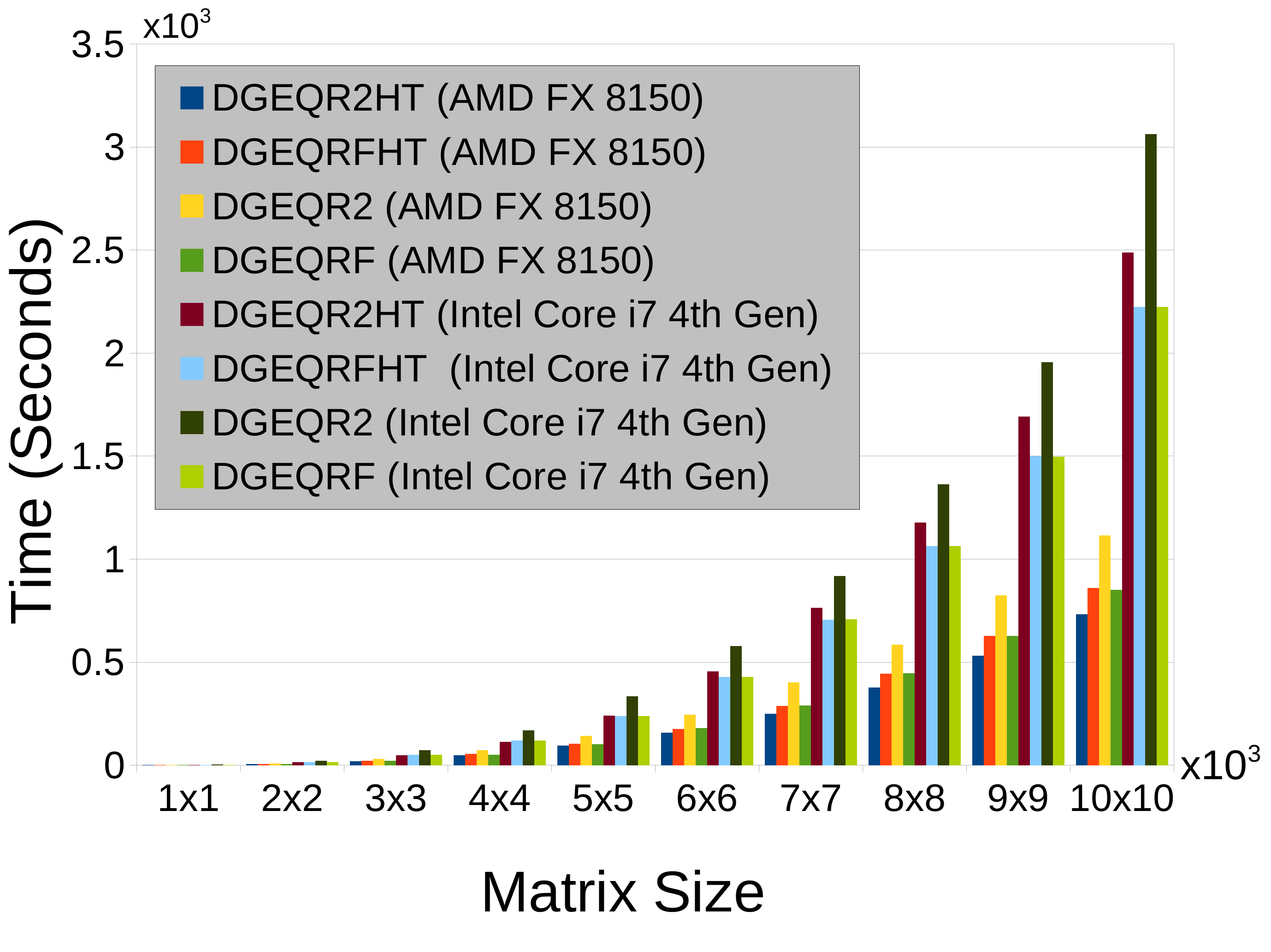}}
\subfigure[Performance Comparison of PLASMA\_DGEQR2, PLASMA\_DGEQRF, DGEQR2HT, and DGEQRFHT on Intel Haswell Mirco-architecture\label{fig:cpi_dgeqr2ht1}]{\includegraphics[scale = 0.21]{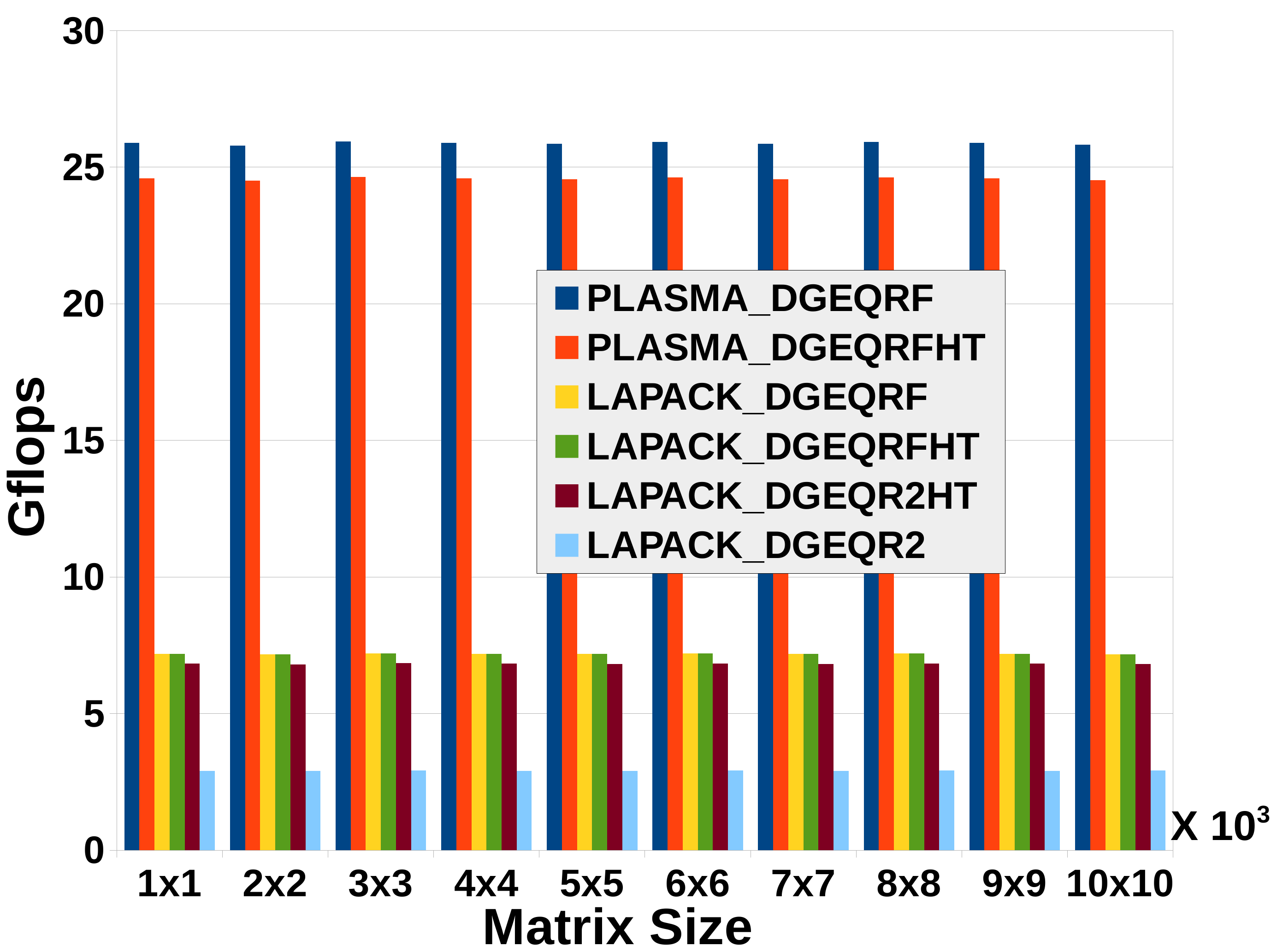}}
\subfigure[Performance Comparison of MAGMA\_DGEQR2, MAGMA\_DGEQRF, MAGMA\_DGEQR2HT, and MAGMA\_DGEQRFHT on Nvidia Tesla C2050\label{fig:cpi_dgeqr2ht2}]{\includegraphics[scale = 0.21]{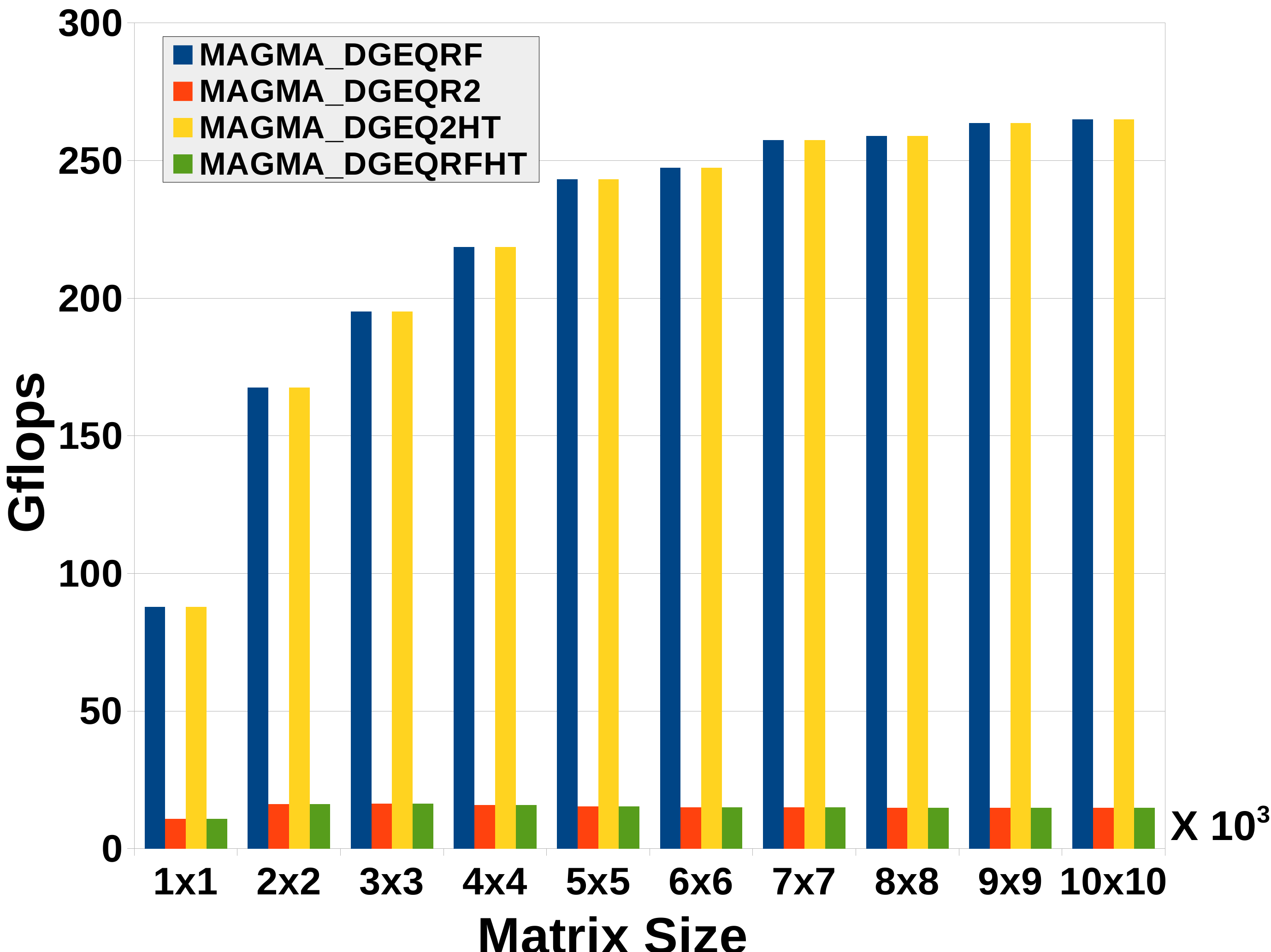}}
\caption{Performance of DGEQR2, DGEQRF, DGEQR2HT, and DGEQRFHT on Commercially Available Micro-architectures}
\label{fig:dgemm_perf}
\end{figure*}

We realize MHT on two different commercially available Intel Haswell and AMD Bulldozer micro-architectures as shown in figure \ref{fig:cpi_dgeqr2ht}. Figure \ref{fig:cpi_dgeqr2ht} also depicts performance of DGEQRFHT where DGEQRFHT is blocked algorithm analogous to DGEQRF in LAPACK. Since PLSAMA is developed using LAPACK as shown in figure \ref{fig:plasma_soft}, we integrate LAPACK\_DGEQRFHT in PLASMA for experiments.  

Pseudo code of LAPACK\_DGEQR2HT is shown in algorithm \ref{algo:dgeqr2ht1}. The routine is implemented as dgeqr2\_ht.f in the source code provided with this exposition. Pseudo code of BLAS\_UPDATE function that is implemented as update1.f is shown in algorithm \ref{algo:update1}. It can be observed in the algorithms \ref{algo:dgeqr2ht1} and \ref{algo:update1} that BLAS\_UPDATE function forms a major computationally intensive part of LAPACK\_DGEQR2HT. BLAS\_UPDATE function becomes part of BLAS that is used inside LAPACK as a basic building block. The routines shown in algorithms \ref{algo:dgeqr2ht1} and \ref{algo:update1} and their wrappers to integrate these routines in LAPACK are supplied with this exposition where directory structure of legacy LAPACK software package is maintained. LAPACK along with Queuing and Runtime for Kernels (QUARK) are used in the PLASMA software stack for multicore realization of LAPACK software package as depicted in figure \ref{fig:plasma_soft} \cite{plasma1}.

\begin{algorithm}
\caption{LAPACK\_DGEQR2HT}
\label{algo:dgeqr2ht1}
\begin{algorithmic}[1]
\State Input: Matrix A of size $M\times N$
\Do
\State Norm = 0, S = 0, B =0
\State Norm = -SIGN(DNRM2(L,X,1), X(1)
\State Beta = (X(1) - Norm)
\State Tau = -Beta/Norm
\State L = M - I + 1
\State UPDATE(L, A(I:M,I), A, LDA, I, M, N, Beta, Norm)
\doWhile(I!=N)
\end{algorithmic}
\end{algorithm}

\begin{algorithm}
\caption{BLAS\_UPDATE}
\label{algo:update1}
\begin{algorithmic}[1]
\Do 
    \State B = A(K,I)*Beta
    \State S = DDOT(L-1, X(2:L),1, A(K+1:M,I),1)
    \State B = B+S
    \State B = B/(Norm*Beta)
    \State A(K,I) = A(K,I) + (Beta*B)
    \State I = K+1
    \Do 
	\State A(J,I) = A(J,I) + A(J,K)*B
	\State J = K+1
    \doWhile(J!=M)
\doWhile(I!=N)
\end{algorithmic}
\end{algorithm}

Similarly, for realization of MHT on GPGPU, we use MAGMA software stack described in figure \ref{fig:magma_soft}. MHT when implemented for GPGPU is shown in algorithm \ref{algo:dgeqr2ht2}.

\begin{algorithm}
\caption{MAGMA\_DGEQR2HT}
\label{algo:dgeqr2ht2}
\begin{algorithmic}[1]
\State Input: Matrix A of size $M\times N$
\State cublasHandle\_t handle
\State magma\_int\_t m, magma\_int\t n
\State magmaFloat\_ptr dA, magma\_int\_t ldda
\State magmaDouble\_ptr dtau, 
\State magmaDouble\_ptr dwork
\State magma\_queue\_t queue
\State magma\_int\_t *info
\State k = min(M,N)
\For{$i=0$ to $k-1$}
  \State MAGMABLAS\_UPDATE(handle, M-i, N-i, dA(i,i), dtau+i, dA(i,i) ldda, dwork, queue)
\EndFor
\end{algorithmic}
\end{algorithm}

\begin{algorithm}
\caption{MAGMABLAS\_UPDATE}
\label{algo:magma1}
\begin{algorithmic}[1]
\If{(m\%BLOCK\_SIZE != 0)}
  \State dim3 grid ((m/BLOCK\_SIZE)+1, 1,1)
  \State threads (BLOCK\_SIZE,1,1)
\ElsIf{(m\%BLOCK\_SIZE = 0)}
  \State dim3 grid ((m/BLOCK\_SIZE), 1,1)
  \State threads (BLOCK\_SIZE,1,1)
\EndIf
  \State cublasDgemv(handle, cublas\_trans\_const(MagmaTrans), M, N, \&alpha, dC, lddc, dv, 1, \&beta, dtau,1)
  \State dtemp$<<<$1, 1, 0, queue $\rightarrow$ cuda\_stream()$>>>$(dC(0,0), dtau, dwork)
  \State dcnst$<<<$grid, threads, 0, queue $\rightarrow$ cuda\_stream()$>>>$(n, dC(0,0), lddc, dtau, dwork)
  \State ddoff$<<<$1,1,0,queue $\rightarrow$ cuda\_stream()$>>>$(dC(0,0), dwork, dwork)
  \State drow1$<<<$grid, threads, 0, queue $\rightarrow$ cuda\_stream(()$>>>$(n, dC(0,0), lddc, dtau, dwork)
  \State dtmup$<<<$n,threads,0,queue $\rightarrow$ cuda\_stream()$>>>$(m, dC(0,0), lddc, dtau, dv)
  \State htcns$<<<$grid,threads,0,queue $\rightarrow$ cuda\_stream()$>>>$(m, dv, dtau, dwork)
\end{algorithmic}
\end{algorithm}

\begin{algorithm}
\caption{dtemp}
\label{algo:magma2}
\begin{algorithmic}[1]
\State Inputs: dot, matrix
\State beta =  sqrt(dot)
\State temp = -copysign(beta, matrix)
\end{algorithmic}
\end{algorithm}

\begin{algorithm}
\caption{dcnst}
\label{algo:magma3}
\begin{algorithmic}[1]
\State Inputs: N, matrix, ldda, temp
\State i = blockIdx.x*blockDim.x + threadIdx.x
\If{(i$<$N)}
  \State dot[i] = MAGMA\_D\_DIV(dot[i], temp[0]*(matrix[0] - temp[0])) - MAGMA\_D\_DIV(matrix[ldda*i], (matrix[0] - temp[0]))
\EndIf
\end{algorithmic}
\end{algorithm}

\begin{algorithm}
\caption{ddiff}
\label{algo:magma4}
\begin{algorithmic}[1]
\State Inputs: matrix, temp
\State diff = matrix - temp
\end{algorithmic}
\end{algorithm}

\begin{algorithm}
\caption{drow1}
\label{algo:magma5}
\begin{algorithmic}[1]
\State Inputs: matrix, ldda, dot, diff 
\State i = blockIdx.x*blockDim.x + threadIdx.x
\If{(i$<$N)}
  \State ltemp =  matrix[ldda*i] + MAGMA\_D\_MUL(dot[i], diff)
  \State matrix[ldda*i] = ltemp
\EndIf
\end{algorithmic}
\end{algorithm}

\begin{algorithm}
\caption{dtmup}
\label{algo:magma6}
\begin{algorithmic}[1]
\State Inputs: M, matrix, ldda, dot, vector 
\State tx = threadIdx.x
\State dot = dot+blockIdx.x
\State matrix = matrix + blockIdx.x*ldda
\If{(blockIdx.x $\ne$ 0)}
  \State tmp = dot[0]
  \For{ j = M-tx-1 to $0$ }
    \State matrix[j] = matrix[j] + tmp*vector[j]
  \EndFor
\EndIf
\end{algorithmic}
\end{algorithm}

\begin{algorithm}
\caption{htcns}
\label{algo:magma7}
\begin{algorithmic}[1]
\State Inputs: M, vector, dtau, diff
\State i = blockIdx.x*blockDim.x + threadIdx.x
\If{(i == 0)}
  \State *dtau = -(diff/vector)
\EndIf
\If{(i$>$0 \&\& i$<$M)}
  \State vector[i] = vector[i]/diff
\EndIf
\end{algorithmic}
\end{algorithm}

It can be observed in the algorithm \ref{algo:dgeqr2ht2} that the computationally intensive part of MAGMA\_DGEQR2HT is MAGMABLAS\_UPDATE function shown in algorithm \ref{algo:magma1}. Similar to BLAS\_UPDATE in LAPACK\_DGEQR2HT where BLAS\_UPDATE is part of BLAS, MAGMABLAS\_UDATE is part of MAGMABLAS. MAGMA\_UPDATE is realized as a series of kernels in CUDA C shown in the algorithms \ref{algo:magma2}, \ref{algo:magma3}, \ref{algo:magma4}, \ref{algo:magma5}, \ref{algo:magma6}, and \ref{algo:magma7}.

It can be observed in figure \ref{fig:cpi_dgeqr2ht} that in AMD Bulldozer micro-architecture LAPACK\_DGEQR2HT performs better than LAPACK\_DGEQR2, LAPACK\_DGEQRF, and LAPACK\_DGEQRFHT. The performance of LAPACK\_DGEQRFHT and LAPACK\_DGEQRF is observed to be same in AMD Bulldozer. In Intel Core i7 $4^{th}$ Gen which is a Haswell micro-architecture, the performance of LAPACK\_DGEQRFHT and LAPACK\_DGEQRF is same. Apart from that LAPACK\_DGEQRFHT and LAPACK\_DGEQRF perform around 10\% better than LAPACK\_DGEQR2HT. When we integrate LAPACK\_DGEQR2HT in PLASMA, the attained performance is depicted in the figure \ref{fig:cpi_dgeqr2ht1} that results in PLASMA\_DGEQRFHT since the trailing matrix update is using LAPACK\_DGEMM along with QUARK. For integration of LAPACK\_DGEQR2HT in PLASMA, we have replaced the instance of LAPACK\_DGEQRF in PLASMA with the instance of LAPACK\_DGEQR2HT. It can be observed in the figure \ref{fig:cpi_dgeqr2ht1} that the performance attained by PLASMA\_DGEQRFHT is 10\% worse than that of performance attained by PLASMA\_DGEQRF. 

Performance of MAGMA\_DGEQR2HT, MAGMA\_DGEQR2, MAGMA\_DGEQRF, and MAGMA\_DGEQRfHT is shown in figure \ref{fig:cpi_dgeqr2ht1}. It can be observed that the performance of MAGMA\_DGQER2 and MAGMA\_DGEQR2HT is almost similar on Nvidia Tesla C2050 while the performance of MAGMA\_DGEQRF and MAGMA\_DGEQRFHT is also similar. Unlike Intel or AMD micro-architectures, performance of MHT in Nvidia is nowhere close to the performance of MAGMA\_DGEQRF )or MAGMA\_DGEQRFHT). This performance figures are not satisfactory since $\beta_{MHT} \ge \beta_{HT}$, and we expect that the performance achieved in LAPACK/PLASMA/MAGMA\_DGEQR2HT is 1.1-1.3x better over LAPACK/PLASMA/MAGMA\_DGEQR2. We also expect that LAPACK/PLASMA/MAGMA\_DGEQR2HT outperforms LAPACK/PLASMA/MAGMA\_DGEQRF. Due to lack of domain customizations for BLAS and LAPACK, we are not able to exploit parallelism that is available in MHT and hence we achieve marginally better performance in multicores while we achieve same performance in GPGPU for realization MHT compared to realization of HT. In this section, we presented modification to classical HT and arrived at MHT where significant improvement in $\beta$ is observed. We further see scope for improvement in realization of MHT where we identify macro operations in MHT and realize them on a specialized RDP that can efficiently execute identified macro operations. 

\section{Custom Realization of Householder Transform and Results}\label{sec:cus}
In this section we present custom realization for HT and show that through algorithm-architecture co-design, performance of the algorithms can be improved significantly. We adopt methodology presented in \cite{exp1}, and \cite{Merc1}, and design a Processing Element (PE) that is efficient in overlapping computations and communication. The PE presented here is also efficient in exploiting Instruction Level Parallelism (ILP) exhibited by BLAS \cite{tpds1}\cite{exp3}. We use this PE as a CFU for REDEFINE for parallel realization to show scalability of algorithms and architecture. 

\subsection{Processing Element Design and Algorithm-architecture Co-design for HT}

\begin{figure*}[!ht]
	\begin{centering}
	\includegraphics[scale=0.20]{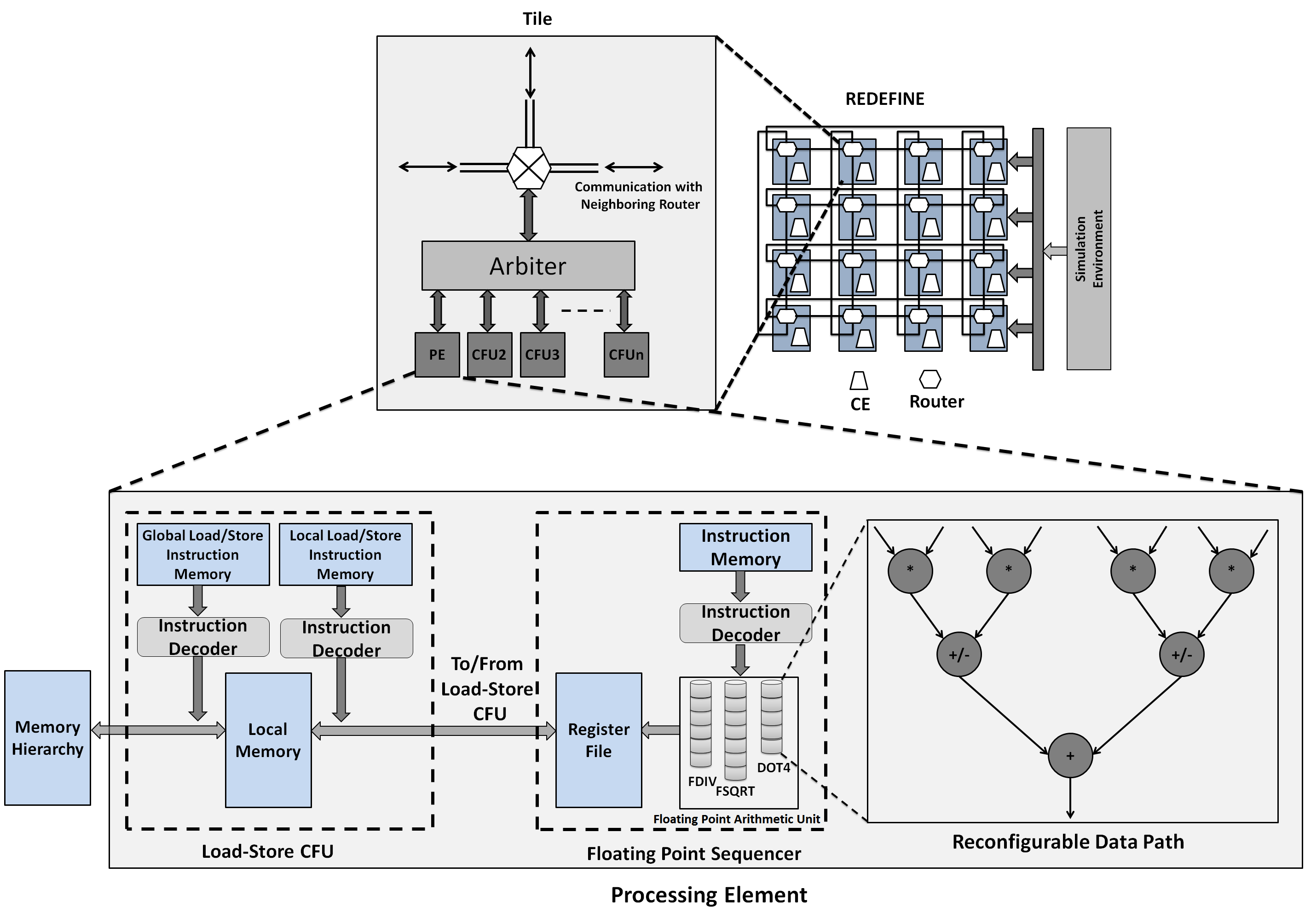}
	\caption{Design of Processing Element}
	\label{fig:pe1}
	\end{centering}
\end{figure*}

Design of PE is depicted in figure \ref{fig:pe1} and it is also shown . It can be observed that PE is separated into two modules: 1) Floating Point Sequencer (FPS), and 2) Load-Store CFU. All the double precision floating point computations are performed in the FPS while Load-Store CFU is responsible for loading/storing data from/to Global Memory (GM) to Local Memory (LM) and LM to Register File, where GM is next level of memory in parallel realization while LM is the private memory of PE, and Register File is small memory of $256$ registers \cite{Merc1}\cite{fpu2}. Operation of PE can be described in following steps: 
\begin{itemize}
 \item {\bf Step 1:} Send a load request to GM for input matrices and store the arriving elements of the input matrices to LM
 \item {\bf Step 2:} Store input matrix elements from LM to Register File 
 \item {\bf Step 3:} Perform computations in FPS
 \item {\bf Step 4:} Store the final/intermediate results from Register File to LM in the Load-Store CFU
 \item {\bf Step 5:} Store final result to GM 
\end{itemize}

\begin{figure}[!ht]
	\begin{centering}
	\includegraphics[scale=0.13]{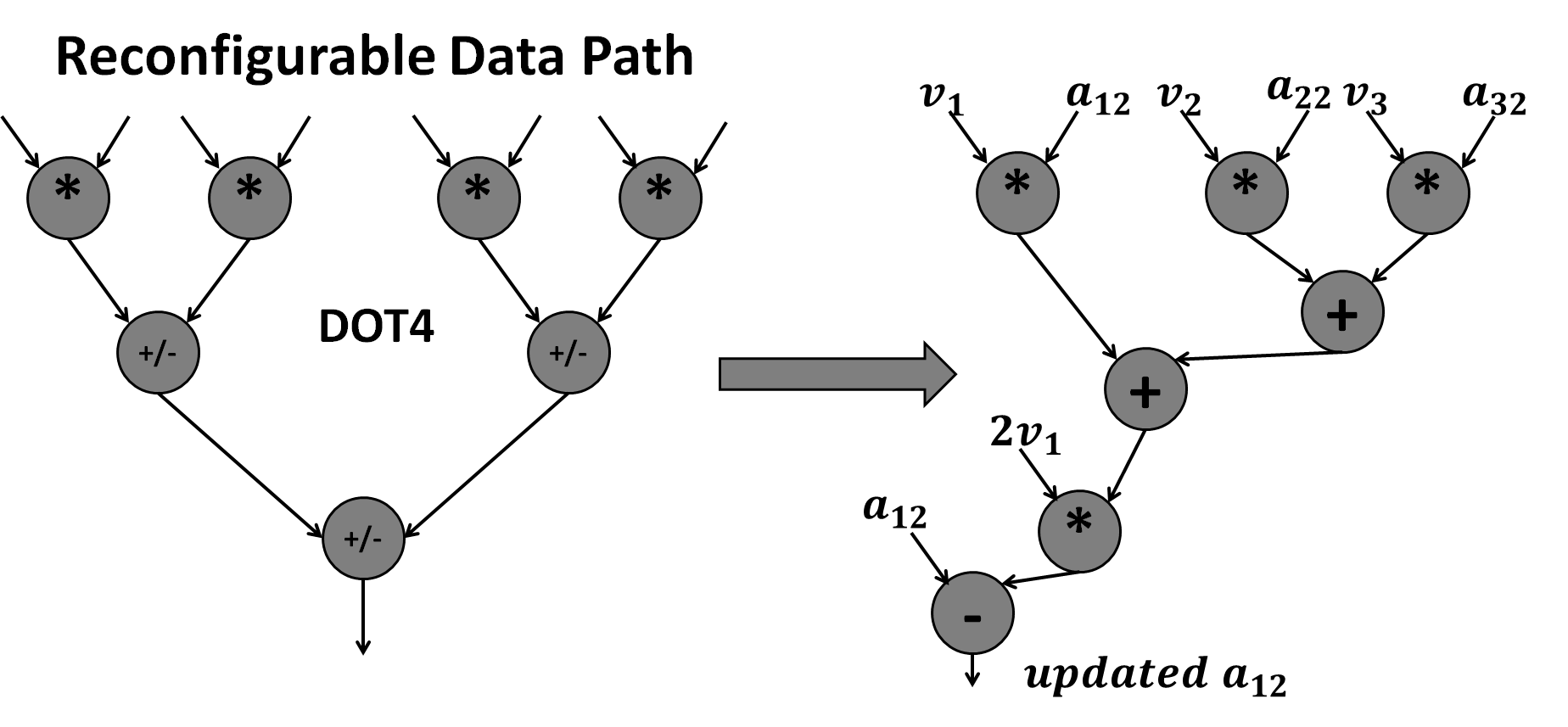}
	\caption{DOT4 and New Configuration of DOT4 for Realization of DGEQR2HT}
	\label{fig:htinst}
	\end{centering}
\end{figure}

In our implementation of DGEQR2, DGEQRF, DGEQR2HT, and DGEQRFHT, we use similar mechanism. FPS has several resources to perform computations. In this exposition, we use carefully designed DOT4, a square root, and a divider for realization of DGEQR2, DGEQRF, DGEQR2HT, and DGEQRFHT routines \cite{fpu2}\cite{fpu3}. Logical place of arithmetic units is shown in the figure \ref{fig:pe1} and structure of DOT4 is shown in figure \ref{fig:htinst}. 

DOT4 can perform inner product of a $4$-element vector. DOT4 is a reconfigurable data-path that can be reconfigured to act as different macro operations encountered in the algorithms \cite{Merc1}. In DGEQR2HT, due to fusion of inner most and intermediate loops, we identify a new macro operation apart from usual macro operations. We explain this with an example of a $3\times 3$ matrix $A = \begin{bmatrix} a_{11} & a_{12} & a_{13}  \\ a_{21} & a_{22} & a_{23} \\ a_{31} & a_{32} & a_{33} \end{bmatrix}$. Applying DGEQR2HT to the matrix $A$ to annihilate $a_{31}$ and $a_{21}$, we compute Householder matrix $P$ and pre-multiply with $A$ as shown in the equation \ref{eqn:ht2}.

\begin{align}\label{eqn:ht2}
	PA = A - 2vv^TA
\end{align} 
where $v = \begin{bmatrix} v_1 \\ v_2 \\ v_3 \end{bmatrix}$ is Householder vector as explained in section \ref{sec:rw}.

\begin{align}
	PA &= \begin{bmatrix} a_{12} & a_{13}  \\ a_{22} & a_{23} \\ a_{32} & a_{33} \end{bmatrix} - 2\begin{bmatrix}v_1 \\ v_2 \\ v_3 \end{bmatrix} \begin{bmatrix}v_1 & v_2 & v_3 \end{bmatrix} \begin{bmatrix}a_{12} & a_{13} \\ a_{22} & a_{23} \\ a_{32} & a_{33}\end{bmatrix} \nonumber \\
	&=  \begin{bmatrix} a_{12} - 2v_1\alpha_1 & a_{13} - 2v_1\alpha_2 \\ a_{22} - 2v_2\alpha_1 & a_{23} - 2v_2\alpha_2 \\ a_{32} - 2v_3\alpha_1 & s_{33} - 2_v3\alpha_2 \end{bmatrix} \label{eqn:upd}
\end{align}
where $\alpha_1 = v_1a_{12} + v_2a_{22} + v_3a_{23}$ and $\alpha_2 = v_1a_{13} + v_2a_{23} + v_3a_{33}$. 

\begin{figure*}
\centering
\subfigure[Speed-up in DGEQR2HT over DGEQRFHt, DGEQR2, and DGEQRF\label{fig:ht_graph_1}]{\includegraphics[scale = 0.21]{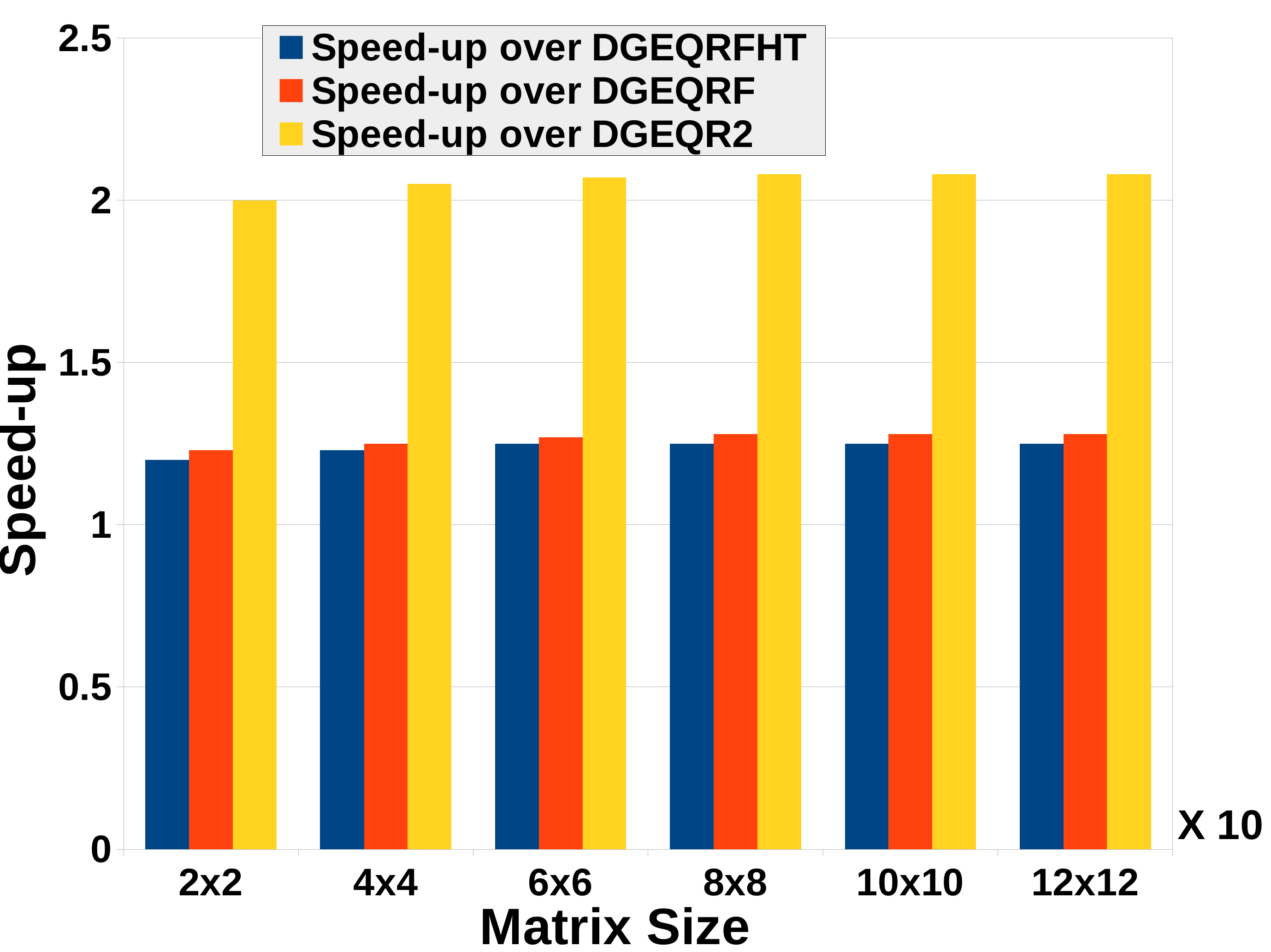}}
\subfigure[Performance Comparison of DGEQR2HT, DGEQRFHT, DGEQR2, and DGEQRF In-terms of Theoretical Peak Performance of the PE and In-terms of the Percentage of Peak Performance Attained by DGEMM in \cite{tpds1}\label{fig:ht_graph_2}]{\includegraphics[scale = 0.21]{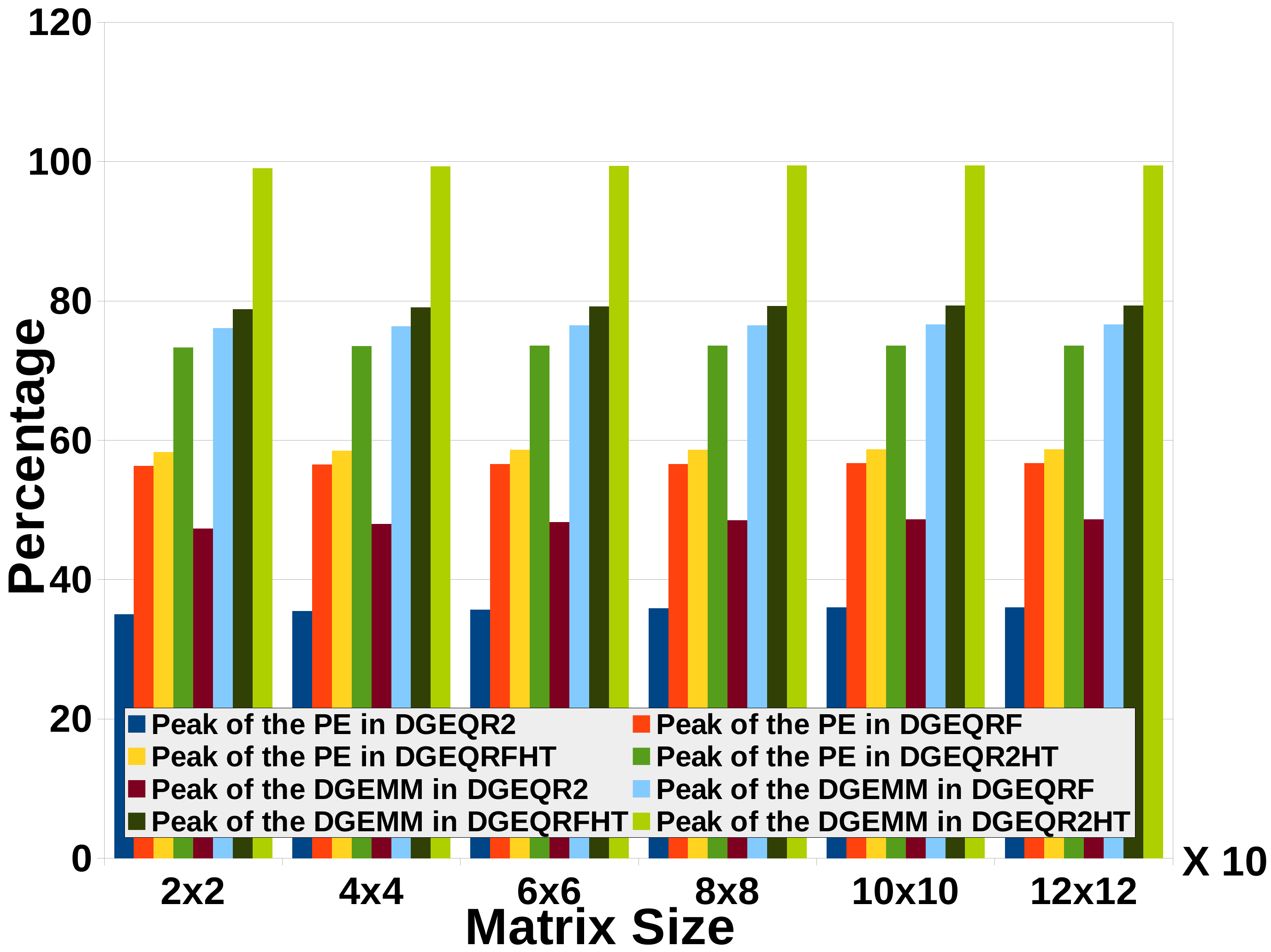}}
\subfigure[Performance Comparison of REDEFINE-PE with Other Platforms\label{fig:ht_graph_3}]{\includegraphics[scale = 0.21]{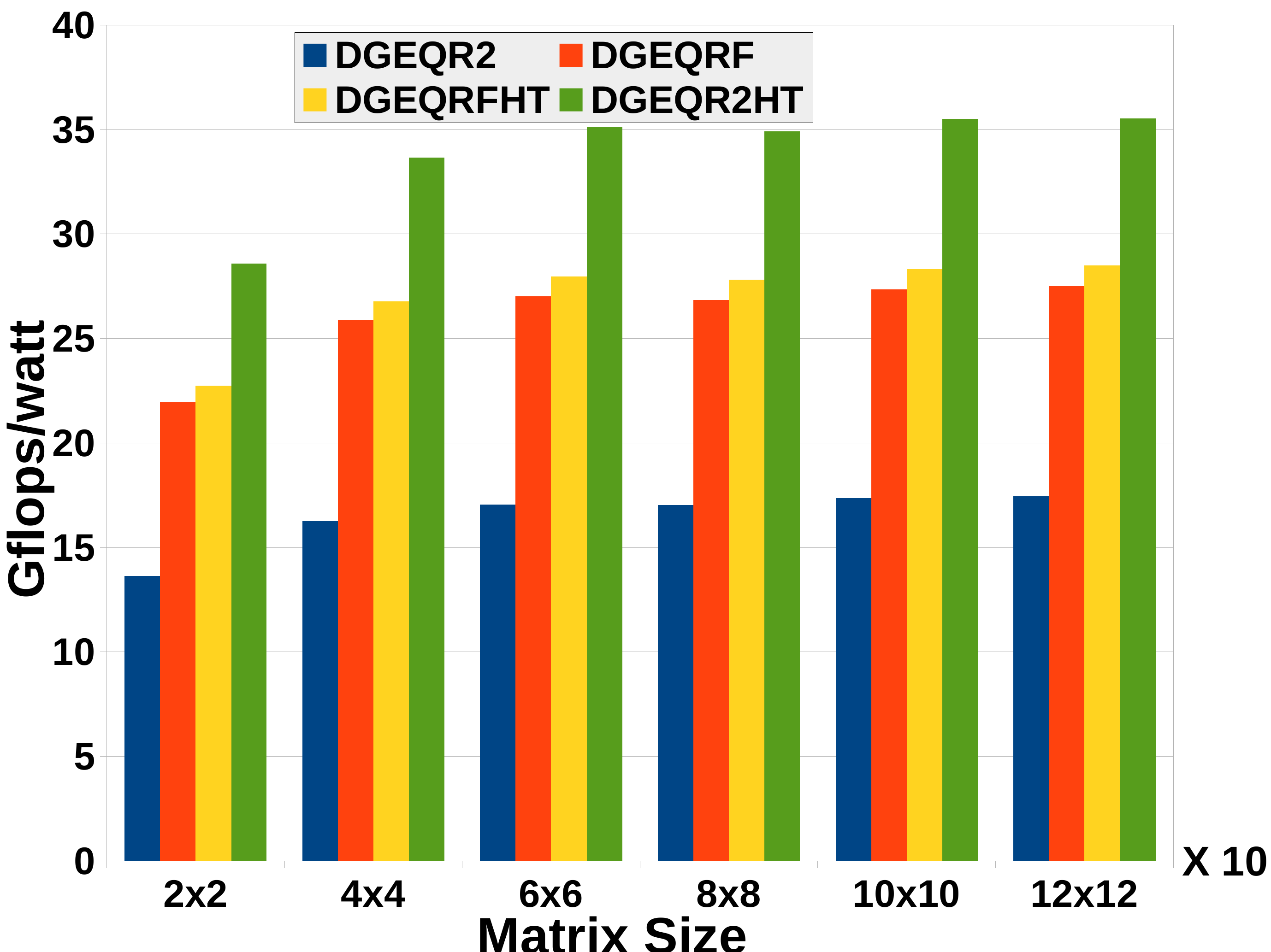}}
\subfigure[Performance Comparison of REDEFINE-PE with Other Platforms\label{fig:ht_graph_4}]{\includegraphics[scale = 0.21]{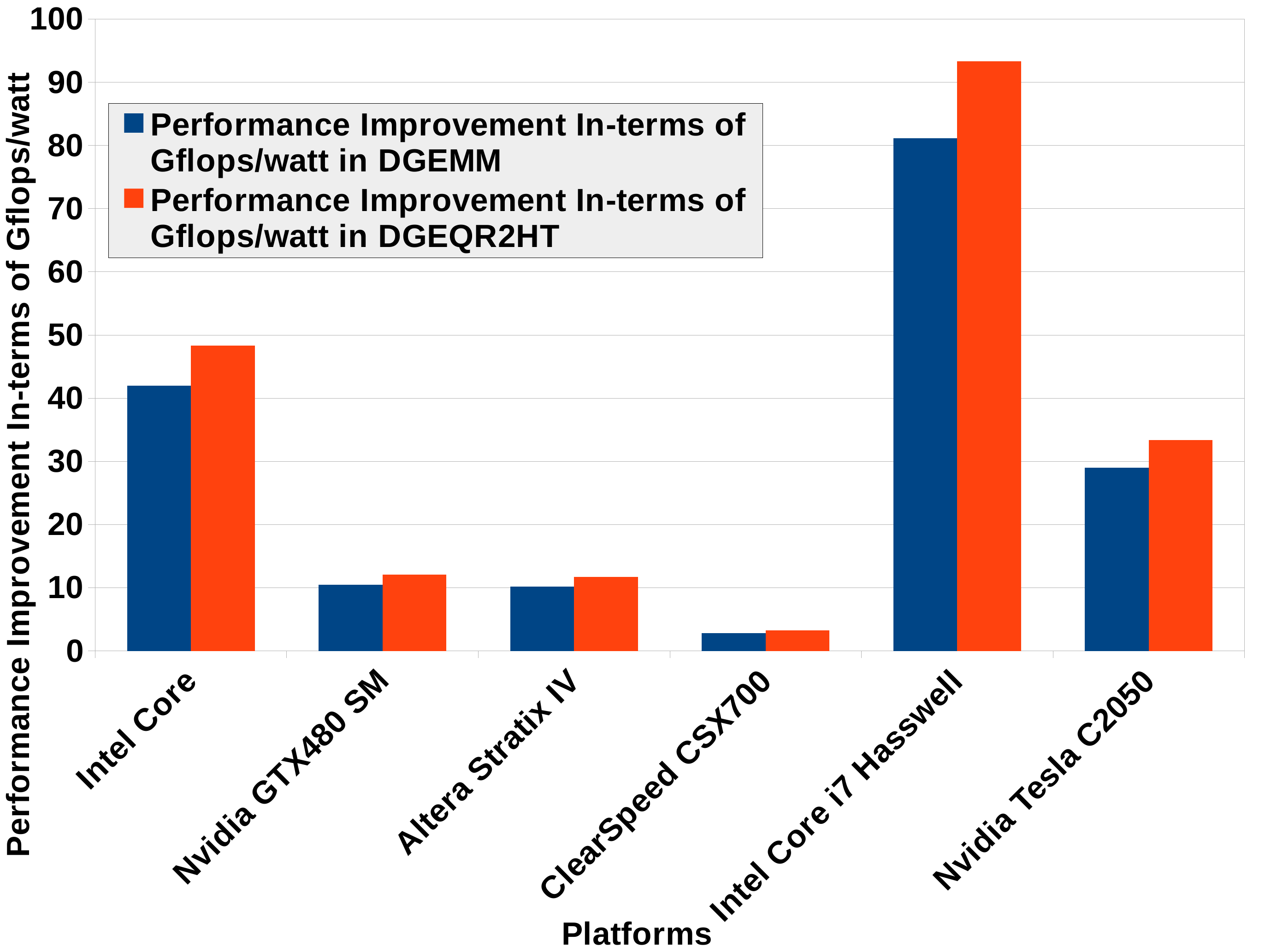}}
\subfigure[Performance Comparison of REDEFINE-PE with Other Platforms\label{fig:ht_graph_5}]{\includegraphics[scale = 0.21]{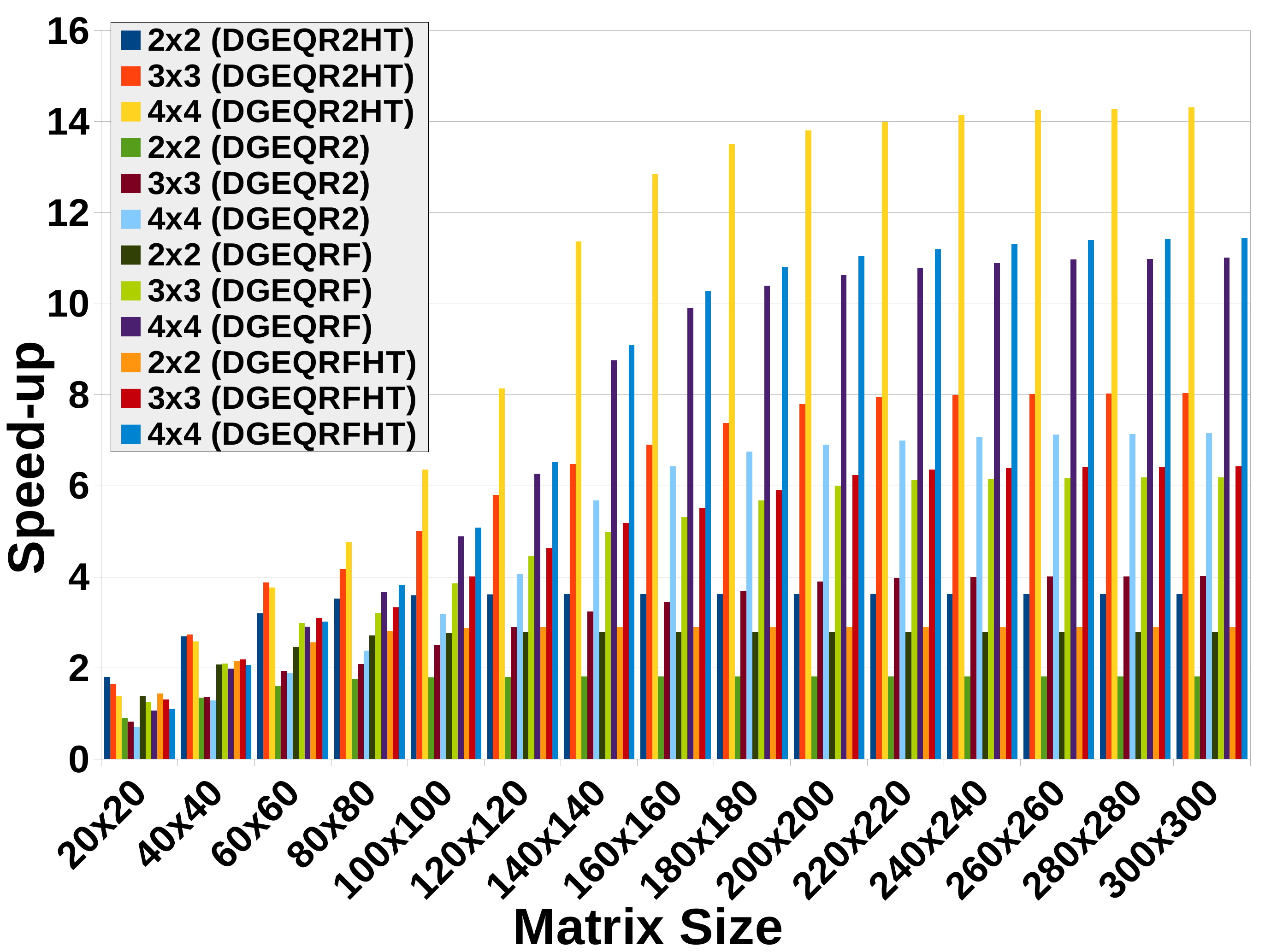}}
\subfigure[Performance Comparison of REDEFINE-PE with Other Platforms\label{fig:ht_graph_6}]{\includegraphics[scale = 0.21]{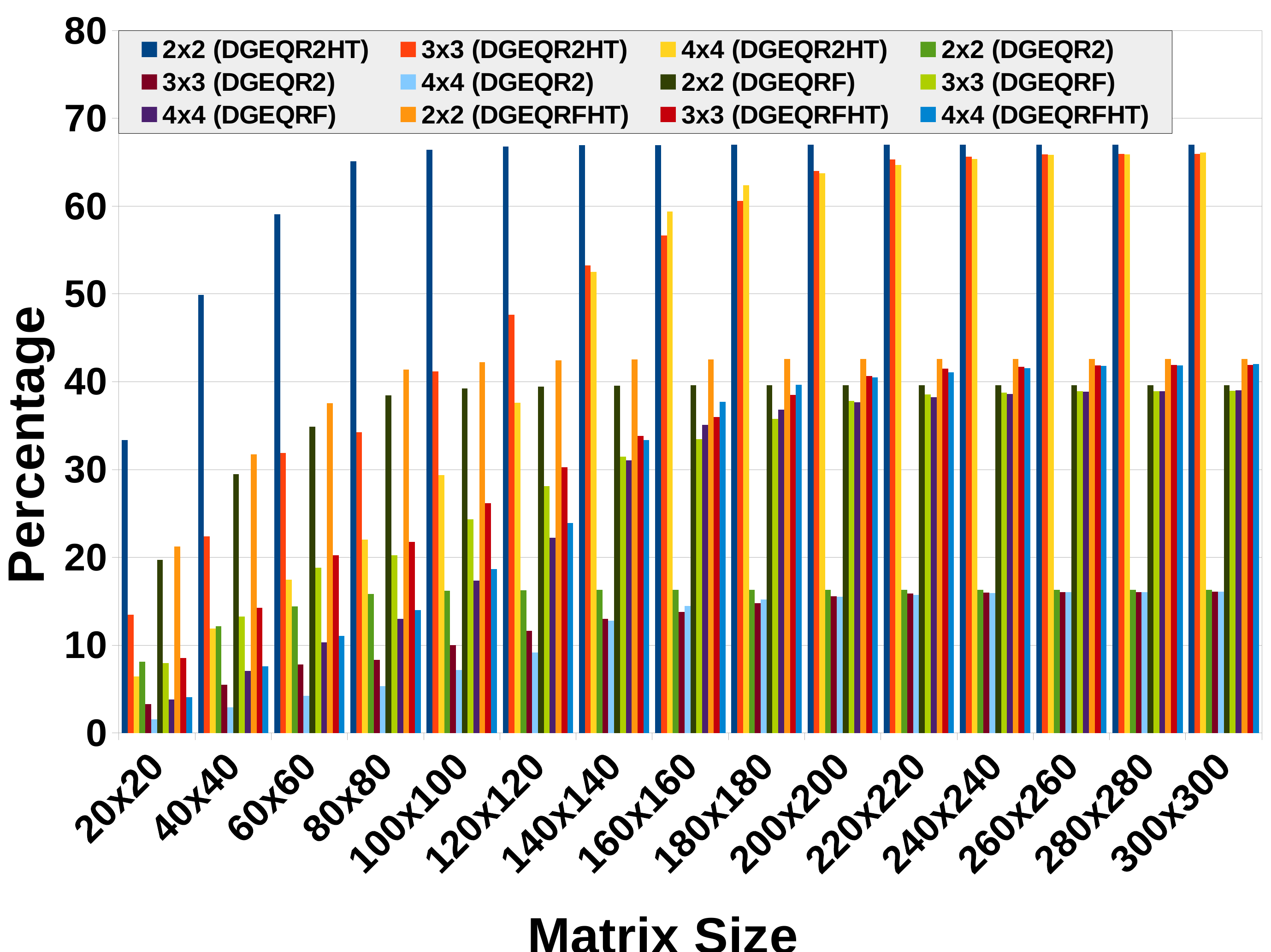}}
\caption{Performance of DGEQR2, DGEQRF, DGEQR2HT, and DGEQRFHT in PE}
\label{fig:dgemm_perf}
\end{figure*}

Taking a close look at the expressions of the updated matrix $PA$ in equation \ref{eqn:upd}, we observe a new macro operation in the expression $a_{12} -2v_2(v_1a_{12} + v_2a_{22} + v_3a_{23})$. For a $3\times 3$ matrix there are $6$ such macro operation encountered as shown in equation \ref{eqn:upd}. We realize this macro operation on DOT4 hardware structure as shown in figure \ref{fig:htinst}. For larger matrices, we break the expressions to fit on the DOT4 hardware structure. Performance after realization of DGEQR2, DGEQRF, DGEQR2HT, and DGEQRFHT is depicted in figure \ref{fig:ht_graph_1}.

It can be observed in figure \ref{fig:ht_graph_1} that DGEQR2HT achieves 2x better performance over DGEQR2, 1.23x over DGEQRF, and 1.23x over DGEQRFHT. Interestingly, DGEQR2HT achieves close to 74\% of the peak performance that is 99.3\% of the performance attained by DGEMM in the PE as shown in figure \ref{fig:ht_graph_2}  In terms of Gflops/watt DGEQR2HT achieves 2x better performance compared to DGEQR2, 1.23x over DGEQRF, and 1.2x over DGEQRFHT. Surprisingly, compared to some of the state-of-the-art realizations of HT based QR factorization, the proposed MHT based QR factorization attains 3-90x better performance which is better than the performance improvement reported for DGEMM in \cite{tpds1}. Such an unexpected result is attained since for the state-of-the-art platforms, the performance attained by LAPACK/PLASMA/MAGMA\_DGEQRF is mostly 80-85\% of the peak performance attained by DGEMM in the platform while in the PE the performance attained by MHT based QR factorization is 99.3\% of the performance attained by DGEMM.

\subsection{Parallel Implementation of HT}
For parallel realization of DGEQR2, DGEQRF, DGEQR2HT, and DGEQRFHT routines, we use simulation environment depicted in figure \ref{fig:pe1}. We attach PE to the Routers in the Tiles of REDEFINE except the last column. In the last column, we attach CFUs with memories that contain same address space. We use this memory as Global Memory (GM). 

\begin{figure*}[!ht]
	\begin{centering}
	\includegraphics[scale=0.30]{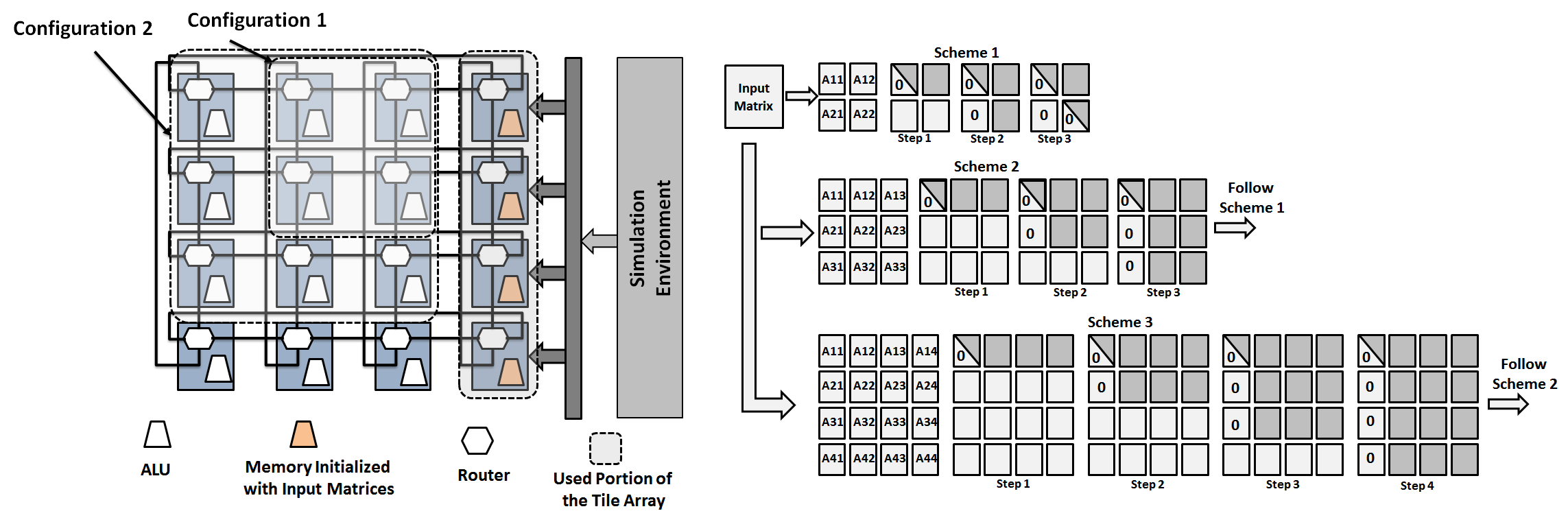}
	\caption{Simulation Environment for Parallel Realization of DGEQR2, DGEQRF, DGEQR2HT, and DGEQRFHT Routines}
	\label{fig:sim_env}
	\end{centering}
\end{figure*}
To show scalability, for experiments, we consider three configurations with different sizes of Tile arrays like $2\times 2$, $3\times 4$, and $4\times 4$. Two configurations are shown in figure \ref{fig:sim_env} namely $Configuration$ $1$ and $Configuration$ $2$ wherein $Configuration$ $1$ is composed of $2\times 2$ Tile array as a fabric for computations and $Configuration$ $2$ is composed of $3\times 3$ Tile array as a fabric for computations. Matrix partitioning schemes for different configurations is also depicted in the figure \ref{fig:sim_env} along with configurations. In the matrix partitioning scheme, we have followed an approach that is used in \cite{plasma1} where input matrix is divided into sub-matrix blocks. For a $K\times K$ fabric of computations and $N\times N$ matrix size, we divide matrix in to the blocks of $\frac{N}{K}\times \frac{N}{K}$ sub-matrices. Since, objective of our experiments is to show scalabillity, we choose $N$ and $K$ such that $N\%K = 0$. Results for parallel implementation of DGEQR2, DGEQRF, DGEQR2HT, and DGEQRFHT routines are depicted in the figure \ref{fig:ht_graph_5}. It can be observed in the figure \ref{fig:ht_graph_6} that the speed-up in parallel realization of DGEQR2, DGEQRF, DGEQR2HT, and DGEQRFHT approaches $K\times K$ when realized using Tile array of size $K\times K$. In figure \ref{fig:ht_graph_5} speed-up attained in parallel realization of any routine is the speed-up over corresponding sequential realizations of the routines. Percentage of theoretical peak performance attained by DGEQR2, DGEQRF, DGEQR2HT, and DGEQRFHT is shown in figure \ref{fig:ht_graph_6}. It can be observed in the figure \ref{fig:ht_graph_6} that DGEQR2HT is capable of attaining 66\% of the theoretical peak of the Tile array utilized for computations while DGEQR2 attains 16\%, DGEQRF attains 39.5\%, and DGEQRFHT attain 42\% of the theoretical peak performance in REDEFINE. DGEQR2HT in REDEFINE clearly outperforms all other routines as depicted in the figure \ref{fig:ht_graph_6} while REDEFINE scales well for DGEQR2, DGEQRF, DGEQR2HT, and DGEQRFHT.

\section{Conclusion}\label{sec:con}
Performance attained by Householder Transform based QR factorization in the state-of-the-art multicore and GPGPUs is usually 80-85\% of the performance attained by General Matrix Multiplication. In this paper, we achieved performance in  Modified Householder Transform similar to the performance of General Matrix Multiplication in terms of Gflops which is contrary to the performance attained in the conventional multicore and GPGPU platforms with the state-of-the-art software packages for Dense Linear Algebra. We moved away from classical approach where optimized Basic Linear Algebra Subprograms are used for realization of Householder Transform based QR factorization and fused inner loops in the Householder Transform based QR factorization based routine (DGEQR2 routine) that resulted in higher degree of parallelism. A simplistic approach for quantification of parallelism was adopted to show 1.3 times higher parallelism in Modified Householder Transform. The classical Householder Transform based QR factorization and Modified Householder Transform based QR factorization along with their optimized block implementations were evaluated on the state-of-the-art multicore and GPGPU where it was observed that the parallelism exhibited by Modified Householder Transform is not fully exploited by these platforms. Design of an existing Processing Element was amended to support the macro operations encountered in Modified Householder Transform by adding a new configuration in the Reconfigurable Data-path in the Processing Element. The approach of realizing macro operations on a Reconfigurable Data-path resulted in 2x performance improvement in Modified Householder Transform over classical Householder Transform in the PE. Realization of Modified Householder Transform could also outperform custom realization of block Householder Transform based QR factorization by 1.2-1.3x. Performance improvement of 3-80x is reported in terms of Gflops/watt over multicore, GPGPU, FPGA, and ClearSpeed CSX700. The performance improvement reported is higher than that of General Matrix Multiplication due to counter-intuitive results obtained in Modified Householder Transform. Finally, it is shown that Processing Element as a Custom Function Unit in REDEFINE results in scalable high performance realization of Householder based and Modified Householder based QR factorization.

\bibliographystyle{IEEEtran}
\bibliography{IEEEabrv,ref}

\begin{IEEEbiography}[{\includegraphics[width=1in,height=1.25in,clip,keepaspectratio]{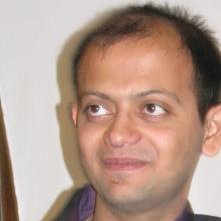}}]{Farhad Merchant}
 Farhad Merchant is a Research Fellow at Hardware and Embedded Systems Lab, School of Computer Science and Engineering, Nanyang Technological University, Singapore. He received his PhD from Computer Aided Design Laboratory, Indian Institute of Science, Bangalore, India. His research interests are algorithm-architecture co-design, computer architecture, reconfigurable computing, development and tuning of high performance software packages
\end{IEEEbiography}
\vspace{-10mm}

\begin{IEEEbiography}[{\includegraphics[width=1in,height=1.25in,clip,keepaspectratio]{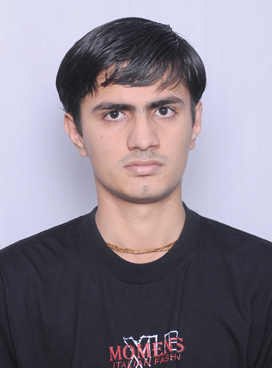}}]{Tarun Vatwani}
Tarun Vatwani is a fresh B.Tech. graduate from Indian Institute of Technology, Jodhpur, India, His research interests are computer architecture, high performance computing, machine learning, performance tuning of different software packages.
\end{IEEEbiography}
\vspace{-10mm}

\begin{IEEEbiography}[{\includegraphics[width=1in,height=1.25in,clip,keepaspectratio]{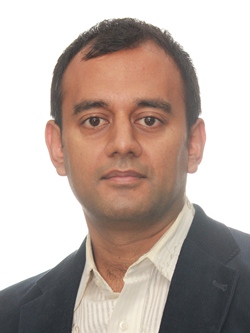}}]{Anupam Chattopadhyay}
 Anupam Chattopadhyay received his B.E. degree from Jadavpur University, India in 2000. He received his MSc. from ALaRI, Switzerland and PhD from RWTH Aachen in 2002 and 2008 respectively. From 2008 to 2009, he worked as a Member of Consulting Staff in CoWare R\&D, Noida, India. From 2010 to 2014, he led the MPSoC Architectures Research Group in RWTH Aachen, Germany as a Junior Professor. Since September, 2014, he is appointed as an assistant Professor in SCE, NTU.
\end{IEEEbiography}
\vspace{-10mm}
\begin{IEEEbiography}[{\includegraphics[width=1in,height=1.25in,clip,keepaspectratio]{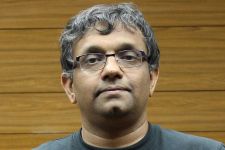}}]{Soumyendu Raha}
 Soumyendu Raha obtained his PhD in Scientific Computation from the University of Minnesota in 2000.
Currently he is a Professor of the Computational and Data Sciences Department at the Indian
Institute of Science in Bangalore, which he joined in 2003, after having worked for IBM for a couple of years.
His research interests are in computational mathematics of dynamical systems, both continuous and combinatorial,
and in co-development and application of computing systems for implementation of computational mathematics algorithms.
\end{IEEEbiography}
\vspace{-10mm}

\begin{IEEEbiography}[{\includegraphics[width=1in,height=1.25in,clip,keepaspectratio]{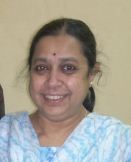}}]
{Ranjani Narayan} Dr. Ranjani Narayan has over 15 years experience at
IISc and 9 years at Hewlett Packard. She has vast work
experience in a variety of fields – computer architecture,
operating systems, and special purpose systems. She
has also worked in the Technical University of Delft, The
Netherlands, and Massachusetts Institute of Technol-
ogy, Cambridge, USA. During her tenure at HP, she
worked on various areas in operating systems and
hardware monitoring and diagnostics systems. She has
numerous research publications.She is currently Chief
Technology Officer at Morphing Machines Pvt. Ltd,
Bangalore, India.
\end{IEEEbiography}

\begin{IEEEbiography}[{\includegraphics[width=1in,height=1.25in,clip,keepaspectratio]{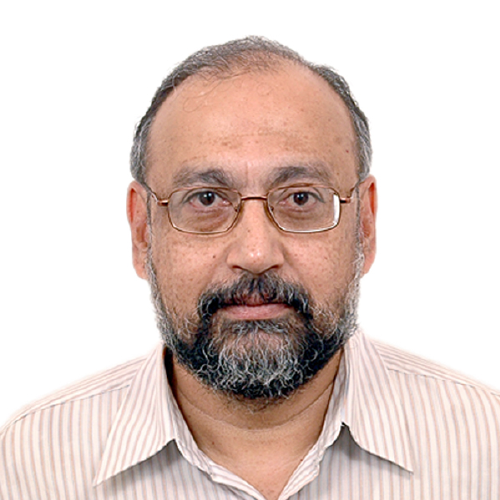}}]{S K Nandy}
S. K. Nandy is a Professor in the Department of Computational and Data Sciences of the Indian Institute of Science, Bangalore.
His research interests are in areas of High Performance Embedded Systems on a  Chip, VLSI architectures for Reconfigurable Systems on Chip,
and Architectures and Compiling Techniques for Heterogeneous Many Core Systems. Nandy received the B.Sc (Hons.) Physics degree from the Indian Institute of Technology, Kharagpur, India, in 1977. He obtained the BE (Hons.) degree
 in Electronics and Communication in 1980, MSc.(Engg.) degree in Computer Science and Engineering in 1986, and the Ph.D. degree in Computer Science
and Engineering in 1989 from the Indian Institute of Science, Bangalore. He has over 170 publications in International Journals, and Proceedings of International Conferences, and 5 patents.
\end{IEEEbiography}

\end{document}